\documentclass[twocolumn,prb,reprint,showpacs,preprintnumbers,amsmath,amssymb,superscriptaddress]{revtex4-1}
\usepackage{graphicx}% Include Fig.~files
\usepackage{dcolumn}% Align table columns on decimal point
\usepackage{bm}% bold math
\usepackage{amsmath}
\usepackage{subfigure}
\usepackage{hyperref}
\usepackage[normalem]{ulem}
\hypersetup{
    colorlinks,
    citecolor=blue,
    filecolor=black,
    linkcolor=blue,
    urlcolor=blue
}
\usepackage{epstopdf}
 \usepackage{relsize}

\newcommand*{\rom}[1]{\expandafter\@slowromancap\romannumeral #1@}
\begin{document}

%\preprint{APS/123-QED}

\title{Role of incoherent scattering on energy filtering in nanostructured  thermoelectric generators.}% Force line breaks with \\

\author{Aniket Singha}

\affiliation{%
Department of Electrical Engineering,\\
Indian Institute of Technology Bombay, Powai, Mumbai-400076, India\\
}%
%\author{Gang Chen}

%\affiliation{
%Department of Mechanical Engineering,\\
%Massachusetts Institute of Technology, 77, Massachusetts Avenue, Cambridge, MA 02139 \\

\author{Bhaskaran Muralidharan}%
\email{bm@ee.iitb.ac.in}
\affiliation{%
Department of Electrical Engineering,\\
Indian Institute of Technology Bombay, Powai, Mumbai-400076, India\\
}%


%\affiliation{
%\affiliation{
%\affiliation{%

\date{\today}% It is always \today, today,

\begin{abstract}
The physics of energy filtering in electronic transport through nanoscale barriers is a fundamental aspect in the context of electronic engineering of nanostructured thermoelectrics.
In the context of thermoelectric generators, it aims to engineer the Seebeck coefficient to favorably increase the power factor and ultimately the power generated.  In this work, we employ the incoherent non-equilibrium Green's function formalism to investigate in detail the physics of energy filtering and how it leads to a direct enhancement in power generation across nanostructured thermoelectrics featuring a single planar energy barrier. In particular, we reinforce that the enhancement in the generated power via energy filtering at a particular operating efficiency is a characteristic of incoherent scattering and is absent in ballistic devices. In such cases, by assuming an energy dependent relaxation time, $\tau(E)=kE^r$, we show that there exists a minimum value $r_{min}$ for which the thermoelectric power generation is enhanced and thereby leading to a degradation in power generation for $r<r_{min}$. For bulk generators, we delve into the details of intermode scattering and show that such scattering processes between electrons in higher energy modes and lower energy modes have a finite contribution to the enhancement in the generated power. We also discuss realistic aspects such as finite width of energy barriers and imperfect energy filtering due to partial reflections. In particular, we show that such imperfect filtering and partial transmission of electrons near the top of the barrier affects the enhancement in the generated power drastically in the high efficiency regime of operation. Analysis of the results obtained in this work should provide general design guidelines for nanostructured enhancement in power generation via energy filtering.
\end{abstract}
\maketitle

%\tableofcontents
\section{Introduction}
The performance of a thermoelectric material is often characterized by the figure of merit ($zT$) defined as 
\[
zT=\frac{S^2\sigma}{\kappa}T
\]
where $S$, $\sigma$ and $\kappa$ are the Seebeck coefficient, the electrical conductivity and the thermal conductivity of the material respectively, and $T$ is the average temperature between the hot and cold contacts. There are typically two distinct approaches followed to facilitate an enhancement in the $zT$ of thermoelectric generators: $(i)$ Lowering the thermal conductivity $\kappa$ and $(ii)$ enhancing the power factor ($S^2\sigma$).  In this context, the approach of nanostructuring via nano-inclusions and interfaces has been successful in suppressing the phonon mediated heat flow due to phonon confinement as well as enhanced phonon scattering \cite{supressk1,supressk2,phonon1,phonon2,phonon3,phonon4,phonon5,phonon6,phonon7,chen3}.  \\
\indent On the other hand, electronic engineering aims at enhancing the electronic figure of merit,  $z_{el}T=\frac{S^2\sigma}{\kappa_{el}}T$, where $\kappa_{el}$ is the electronic thermal conductivity.  Manipulating the electronic density of states (DOS) \cite{hicks1,hicks2,dresselhaus,goldsmid,response1,response2,extra1,extra2,mypaper,boyu} in both low-dimensional systems and bulk systems with nanoinclusions and energy barriers is a topic of intense and active research \cite{response1,response2,response3,response4,mypaper,hicks1,hicks2,dresselhaus,murphy,poudel,snyder,heremans,Mona_Z}. In this aspect, theoretical investigations on bulk materials with semiconducting/metallic inclusions and energy barriers have demonstrated a power factor enhancement \cite{theory1,theory2,theory3,vashaee,chen_ref,vining1,vining2}. This enhancement is largely attributed to the filtering of lower energy electrons due to the interface potentials \cite{theory1,theory2,theory3,Mona_Z,vashaee}. \\
\indent From a fundamental thermodynamic stand point however, analysis of nanoscale thermoelectric devices solely on the basis of the figure of merit picture is somewhat inadequate \cite{jordan1,jordan2,sothmann,bm,whitney,whitney2}. The figure of merit $zT$, albeit a handy metric, is typically valid only in the linear response regime and most importantly cannot facilitate a clear understanding of the physics of heat flow in the nanoscale. Particularly, it only relates to the maximum efficiency point and does not take in to consideration other operating points of importance in the thermoelectric generator set up. This is specifically relevant to the analysis of power generation in the context of nanoscale thermoelectric generators under varying operating conditions . Therefore a non-equilibrium analysis of power generation at a given thermodynamic efficiency and operating point \cite{whitney,whitney2,sothmann,bm,agarwal} is essential for a thorough analysis of design strategies \cite{nakpathomkun,jordan2,agarwal,bm,zimb,leij}.\\
\indent In this paper, we employ the non-equilibrium Green's function (NEGF) formalism with the inclusion of incoherent scattering \cite{Lake_Datta} to delve into the physics of energy filtering in nanowires and bulk generators with planar nanoscale energy barriers. We point to conditions for optimum energy filtering and the enhancement of the generated thermoelectric power. It is shown that such an enhancement of generated power due to energy filtering is characteristic to systems dominated by incoherent electron scattering processes and that coherent scattering in general cannot contribute likewise.  We also discuss in detail other realistic factors such as imperfect filtering, different scattering mechanisms and realistic barrier parameters that contribute to the power enhancement in nano structures and bulk generators.\\
\indent This paper is organized as follows. In Sec. \ref{filtering}, we discuss the basic concepts related to energy filtering and how it is related to the enhancement of generated power in thermoelectric generators. We define a relevant metric, the \emph{filtering coefficient}, for a quantitative analysis of the enhancement in power generation due to energy filtering. In Sec. \ref{model}, we briefly discuss briefly the details of the models employed while a detailed exposition of the formalism used is carried out in the appendices. Sections \ref{coherent} and \ref{incoherent} deal with energy filtering and power generation in devices dominated by coherent and incoherent scattering respectively. Next, in Secs. \ref{length} and \ref{order}, we discuss the dependence of the filtering coefficient on the length of the device and the order of the scattering mechanism. In doing so, we delve into relevant details such as, the contribution of intermode coupling in Sec. \ref{intermode} to the generated power in bulk thermoelectric generators. We then incite a brief discussion on perfect versus imperfect filtering in systems dominated by incoherent scattering in Secs. \ref{imperfect} and \ref{combo}, and draw general conclusions in Sec. \ref{conc} .
\section{Energy filtering in semiconductors}\label{filtering}
We start with a basic exposition on the physics of energy filtering from a linear response point of view. While dealing with nanostructures, it is more convenient to work with extensive linear response properties such as conductance rather than intensive properties such as conductivity \cite{nakpathomkun,LNE}. 
For thermoelectric power generation, a measure of the maximum generated power per unit area in the linear response regime is represented by $S^2G$, where $S$ is the Seebeck coefficient and $G$ is the conductance of the device per unit area. The connection with the actual power generated may be easily deduced from the linear response expansions of the heat and electric currents with respect to the applied voltage and temperature gradient \cite{goldsmid,LNE}. \\
\indent In the linear response limit, the Seebeck coefficient is given by 
\begin{equation}
S=-\frac{k_B}{q}\left[\frac{E_C-E_F}{k_BT}+\frac{\int \frac{E-E_C}{k_BT}G (E) dE}{\int G(E)dE}\right]
\label{eq:seebeck}
\end{equation}
while the conductance is given by 
\begin{equation}
G=\int G(E)dE
\end{equation}
In the above equation, $E_C$ denotes the energy at the conduction band minima, $E_F$ denote the Fermi energy. For ballistic devices, in the linear response limit,  $G(E)$ is given by 
\begin{equation}
G(E)=\frac{1}{A}\frac{e^2}{h} \sum_m T_m(E)\{-\frac{\partial f}{\partial E}\} dE,
\label{eq:ballistic_conductivity}
\end{equation}
where $A$ is the cross-sectional area of the device and $T_m(E)$ is the transmission function for $m^{th}$ mode at energy $E$ \cite{nakpathomkun}. The summation in \eqref{eq:ballistic_conductivity} includes all the modes contributing to conductance of the device. For  devices dominated by incoherent scattering, the conductance  per unit area is given by 
\begin{equation}
\frac{1}{G(E)}=\int \frac{1}{\sigma(z,E)}dz,
\end{equation}
with $z$ being the transport direction and $\sigma(z,E)$ being the energy resolved electrical conductivity of the device at point $z$ and energy $E$  given by \cite{ashcroft}:
\begin{equation}
\sigma(z,E)=v^2_z(z,E) \tau(z,E) D(z,E) \{-\frac{\partial f(z,E)}{\partial E}\},
\label{diffusive_conductivity}
\end{equation}
in the diffusive limit, where $v(z,E)$, $D(z,E)$, $\tau(z,E)$ and $f(z,E)$ represent the band velocity, the local density of states as a function of energy, the position and energy resolved relaxation time and the carrier distribution as a function of energy respectively.
The above equation assumes a quasi-equilibrium distribution for electrons throughout the entire device.\\
%In the linear response limit, the Seebeck co-efficient is given by 
%\begin{equation}
%S=\frac{\Pi}{T}
%\label{eq:average_energy}
%\end{equation}
%where $\Pi$ is known as the Peltier coefficient and is a measure of the average energy current flowing through the device.
\indent The principal objective of energy filtering that is widely discussed in literature \cite{theory1,theory2,theory3,kim1,kim2} is to enhance the Seebeck coefficient by allowing only the high energy electrons to contribute to the conductance thereby increasing the average energy carried by the current. However, such a filtering of electrons decreases the overall conductivity and this trade-off needs to be delved into further. In general, energy filtering is achieved with the help of metallic nano inclusions \cite{theory2,nanoinclusion1,nanoinclusion2,nanoinclusion3,nanoinclusion4,nanoinclusion5,nanoinclusion6}  or potential energy barriers \cite{theory1,theory3,kim1,kim2} to inhibit the lower energy electrons from participating in conduction while still allowing the higher energy electrons to contribute to the total current. Unlike metals, semiconductor devices have attracted recent attention in the context of thermoelectric energy filtering because the electrochemical potential  can be modulated easily via a change in doping concentration. In this paper, we study energy filtering from the perspective of the order of electronic scattering mechanism, with a relaxation time given by $\tau(E)=k_1E^{r}$, with a smaller exponent denoting a higher order scattering process. We show that there is a minimum value of $r$ ($r>r_{min}$) beyond which energy filtering enhances thermoelectric generation. Instead of studying the change in $S$ and $G$ separately due to energy filtering, we study the relative enhancement in power generation via a single parameter $r$. The whole concept then reduces to the fact that the dependence of $\tau(E)$ on energy determines the quantitative enhancement in the generated power with a relatively  higher value of $r$ contributing to a relatively larger enhancement in the generated power due to energy filtering.  \\ 
\indent In this work, we follow the non-equilibrium approach to deduce various currents from the NEGF formalism to be introduced shortly, following which a direct calculation of power and efficiency is performed. The power $(P)$ and the efficiency ($\eta$) in the case of a thermoelectric generator with an applied temperature difference across two contacts can be defined as:
\begin{equation}
P=I_C \times V
\end{equation} 
\begin{equation}
\eta=\frac{P}{I_Q},
\end{equation}
where $I_C$ is the charge current, $I_Q$ is the heat current at the hot contact due to electronic heat conductivity and $V$ is the applied voltage assuming a voltage controlled set up described in recent literature\cite{leij,sothmann,bm,nakpathomkun,jordan2,mypaper,akshay} . \\
\indent In order to assess the efficacy of electron filtering across the barrier, we now define a metric, \emph{filtering coefficient ($\lambda$)} as 
\begin{equation}
\lambda(\eta)=\frac{P_f(\eta)}{P_{nf}(\eta)},
\label{eq:fc}
\end{equation}
where $P_f(\eta)$ and $P_{nf}(\eta)$ are the maximum power densities at efficiency $\eta$ with and without energy filtering respectively. For thermoelectric generators, $P_f(\eta)$ and $P_{nf}(\eta)$ are taken along the operating line of the device \cite{mypaper}. The exact value of $\lambda(\eta)$ depends on the material parameters of the device and the shape of the energy barrier used. Our intention in this paper is to give a general idea on the enhancement of $\lambda(\eta)$ as a result of energy filtering.
\section{Transport Formulation}\label{model}
\begin{figure}[]
\subfigure[]{\includegraphics[scale=.04]{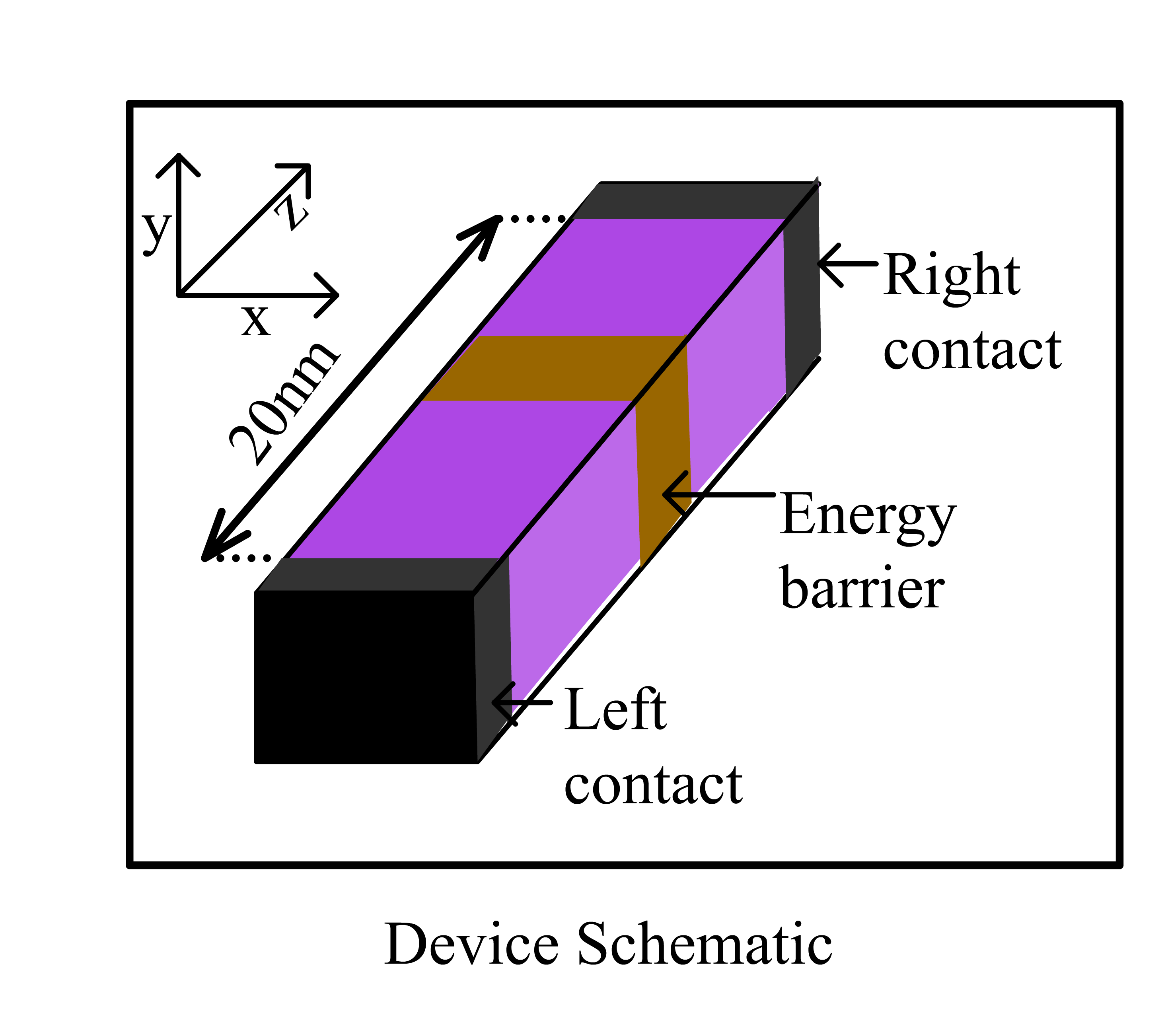}
}\subfigure[]{\includegraphics[scale=.18]{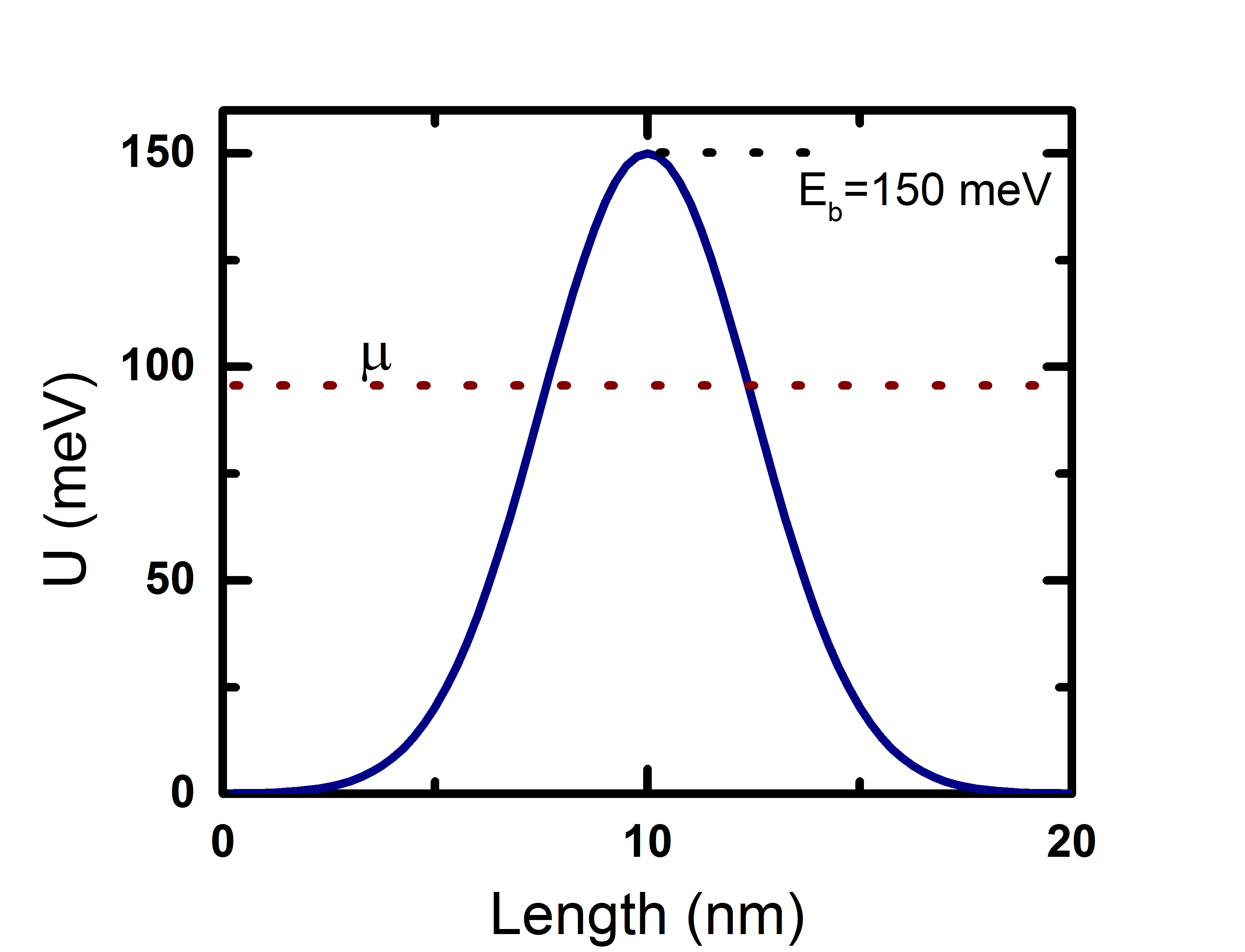}}
\caption{ Device Schematics. (a) The device used for simulation, with a device region length of $=20nm$. (b) The band profile of the device embedded with a Gaussian energy barrier of height $E_b=150meV$ and $\sigma_w=2.7nm$. The brown dotted line shows equilibrium electrochemical potential of the device for the case $E_b-\mu_0=2k_BT$.  }
\label{fig:schematic}
\end{figure}

We consider thermoelectric generators in which the active regions are smaller than the energy relaxation lengths \cite{kim1} such that the energy current is almost constant throughout the device region. For the purpose of the simulations, we use the band parameters of the $\Delta_2$ valley of lightly doped silicon \cite{book1} with  a longitudinal effective mass, $m_l=0.98m_0$, and a transverse effective mass, $m_t=0.2m_0$. A schematic of the generic device structure used is shown in Fig. \ref{fig:schematic}(a) along with the band diagram schematic of an embedded Gaussian energy barrier of height $E_b=150meV$ and $\sigma_w=2.7nm$, depicted in Fig. \ref{fig:schematic} (b).
\[
U=E_b exp\left(-\frac{x^2}{2\sigma_w^2}\right),
\]
where the mid-point of the device region is taken as $x=0$. The transverse geometries of the device region considered here include bulk, where the transverse extent is infinite and nanowires, where the transverse extent consists of only one  sub-band.\\
\indent Such barriers can be fabricated in a $Si/Si_xGe_{1-x}$ heterojunction by making an appropriate choice of $x$ and the doping concentration. A brief discussion on the spatial variation of bands, bandstructure and band offsets with change in doping concentration and manipulation of relative concentrations of $Si$ and $Ge$ in $Si_xGe_{1-x}$ heterostructures is carried out in Refs. \cite{sige1,sige2,sige3,sige4}.  In fabricated heterojunctions, the effect of inhomogeneous doping, disorder and unequal work functions smoothens out thin energy barriers \cite{theory3,kim1,kim2} into  Gaussian/bell-shaped profiles. For such barriers, electron filtering is imperfect and hence electrons with energy near $E_b$ would be partially transmitted  due to coherent reflection from the top of the barrier. We will show that such an imperfect energy filtering affects the performance of the device drastically in the high efficiency regime of operation.  %Figure compares the transmission probability of electrons in the energy range between for a clean device and a device with Gaussian energy barrier $E_b$. 
A more rigorous approach would be to self-consistently solve for the potential and charge density along the device using the information on band-offsets and doping profile along the device.  However, we do not expect significant deviations from the conclusions derived in this paper because our results are strongly dependent on the type of scattering mechanism, which is the principal focus of this work. In addition, we are interested in the conditions under which energy filtering in thermoelectric generators enhance the generated power at a given efficiency. Since thermal conductivity due to phonon doesn't vary significantly with barrier height and width, it is logical to simplify our calculations by neglecting the degradation in efficiency due to phonon heat conductivity.\\
\indent In order to perform the transport calculations to be presented, we employ the NEGF transport formalism with the self-consistent Born approximation \cite{dattabook,Lake_Datta} to incorporate scattering in the device region. We start with the single particle Green's function $G(\overrightarrow{k_{m}},E_z)$, for each transverse sub-band $m$ \cite{dattabook}, evaluated from the device Hamiltonian matrix [H] given by:
\begin{gather}\label{eq:negf_main}   
G(\overrightarrow{k_{m}},E_z)=[E_ZI-H-U-\Sigma(\overrightarrow{k_{m}},E_z)]^{-1} ,\nonumber \\
\Sigma(\overrightarrow{k_{m}},E_z)=\Sigma_L(\overrightarrow{k_{m}},E_z)+\Sigma_R(\overrightarrow{k_{m}},E_z)+\Sigma_s(\overrightarrow{k_{m}},E_z) \nonumber \\
\end{gather}
where, $[H]=[H_0]+U$, with $[H_0]$ being the device tight-binding Hamiltonian matrix \cite{dattabook} and $I$ being the identity matrix of the same dimension as the Hamiltonian.  $E_z$ is the free variable denoting  the energy of the $z$ extent of the electron wavefunction and  the spatial variation in the conduction band minimum is described by the matrix $U$. The vector $\overrightarrow{k_{m}}$ represents the wavevector of the electron in the transverse direction for the $m^{th}$ mode. The net scattering self energy matrix $[\Sigma(\overrightarrow{k_{m}},E_z)]$ includes that due to the scattering of the electronic wavefunctions from the contacts into the device region, denoted by $\Sigma_L(\overrightarrow{k_{m}},E_z)+\Sigma_R(\overrightarrow{k_{m}},E_z)$ as well as the scattering of electronic wavefunctions inside the device due to phonons and non-idealities, denoted by  $\Sigma_s(\overrightarrow{k_{m}},E_z) $. Under the self consistent Born approximation scheme, the electronic scattering rates due to contact couplings and electron-phonon interactions are given by the in-scattering and out-scattering functions, $[\Sigma^{in}]$ and $[\Sigma^{out}]$, as detailed in Appendix \ref{appendix1}, specifically in \eqref{eq:sig}, \eqref{eq:sig1} and \eqref{eq:sigma}. The calculation of the in-scattering and the out-scattering functions involve a self consistent procedure, detailed in Appendix \ref{appendix1}, with the electron and the hole density operators $G^n(\overrightarrow{k_{m}},E_z)$, $G^p(\overrightarrow{k_{m}},E_z)$ defined as
\begin{eqnarray}
G^n(\overrightarrow{k_{m}},E_z)=G(\overrightarrow{k_{m}},E_z)\Sigma^{in}(\overrightarrow{k_{m}},E_z)G^{\dagger}(\overrightarrow{k_{m}},E_z), \nonumber \\
G^p(\overrightarrow{k_{m}},E_z)=G(\overrightarrow{k_{m}},E_z)\Sigma^{out}(\overrightarrow{k_{m}},E_z)G^{\dagger}(\overrightarrow{k_{m}},E_z). \nonumber \\
\end{eqnarray}
Upon convergence of the self-consistent quantities, the charge and heat currents are evaluated in the lattice basis as:
\begin{eqnarray}
I^{j\rightarrow j+1}=\underset{k_m}{\sum}\frac{i}{\pi \hslash} \int[H_{j+1,j}(E_z)G^n_{j,j+1}(\overrightarrow{k_{m}},E_z) \nonumber \\
-G^n_{j+1,j}(\overrightarrow{k_{m}},E_z)H_{j,j+1}(E_z)]dE_z \nonumber \\
\label{eq:currentnegf}
\end{eqnarray}
\begin{eqnarray}
I_Q^{j\rightarrow j+1}=\underset{k_m}{\sum}\frac{i}{\pi \hslash}   \int (E_z+E_m-\mu_H)[H_{j+1,j}(E_z)\nonumber \\
G^n_{j,j+1}(\overrightarrow{k_{m}},E_z) 
-G^n_{j+1,j}(\overrightarrow{k_{m}},E_z)H_{j,j+1}(E_z)]dE_z, \nonumber \\
\label{eq:heatcurrentnegf}
\end{eqnarray}
where $\mu_H$ is the electrochemical potential of the hot contact, $M_{i,j}$ is a generic matrix element of the concerned operator between two lattice points $i$ and $j$. In the tight-binding scheme used here, we only consider the next nearest neighbor such that $j=i \pm 1$. The heat current $I_Q$ is evaluated at the hot contact and will be of relevance in our calculations. \\
\indent In the following calculations, the temperatures of the hot and cold contacts, labeled $H$ and $C$, are assumed to be $330K$ and $300K$ respectively. Without loss of generality, we set the left (right) contact, $L(R)$, to be the hot (cold), $H(C)$ contact. The power density versus efficiency calculations are carried out  by varying the bias voltage to emulate an external current flow. By varying the bias voltage continuously, the power density and the efficiency $\eta$ are calculated to generate a set of points on the $\eta-P$ plane for a particular position of the equilibrium electrochemical potential $\mu_0$. We assume that both the contacts are symmetrically coupled and that the potential drop is linear across the energy barriers. Assuming
quasi-equilibrium electronic distribution at the contacts, their respective electrochemical potentials are given by $\mu_{H/C}=\mu_{0}\pm V/2$. One can then say that $V$ is the potential drop across the thermoelectric generator due to current flow through an external passive circuit element. 
\section{Results}
\subsection{Energy filtering with coherent scattering}\label{coherent}
\begin{figure}[]
\subfigure[]{\hspace{-.5cm}\includegraphics[scale=.18]{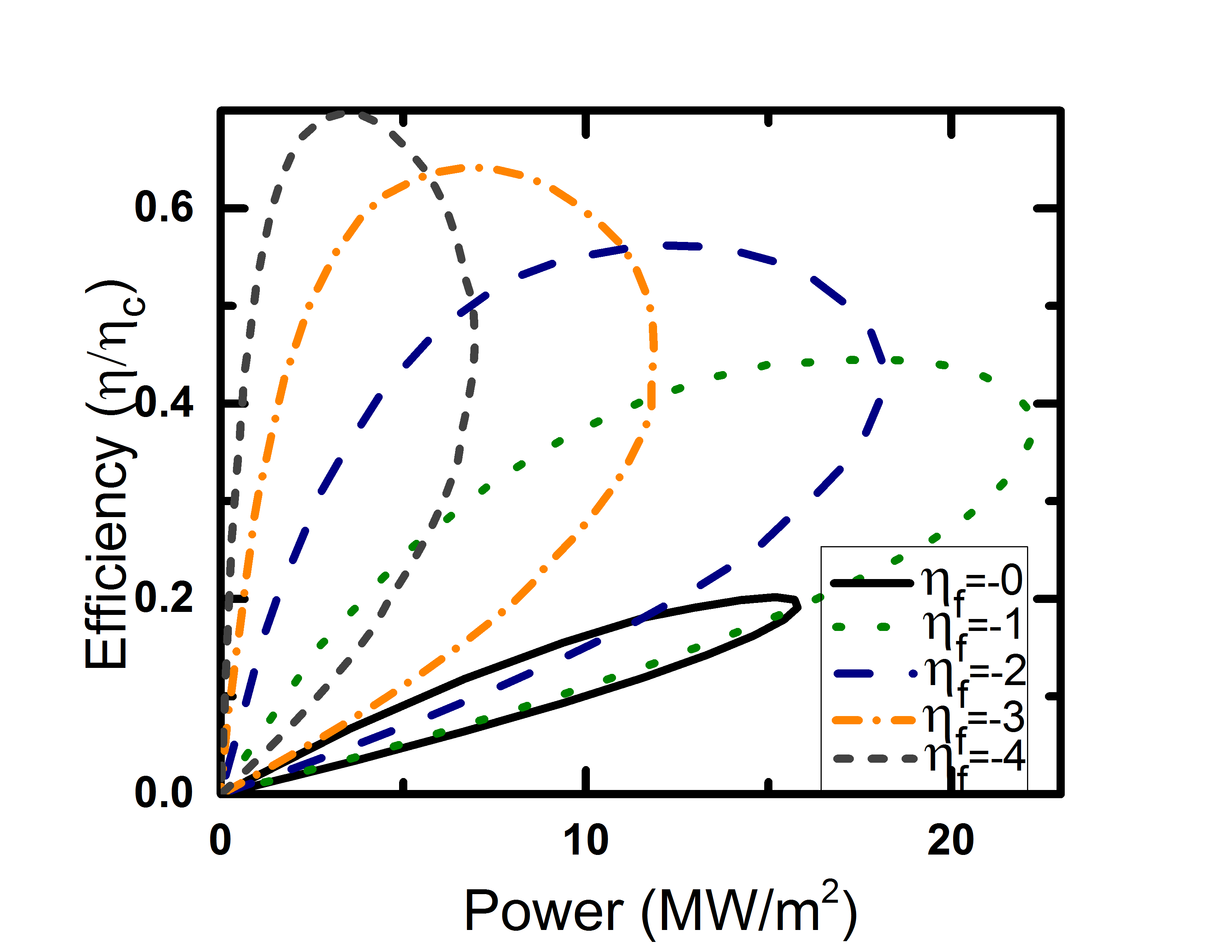}
}\subfigure[]{\hspace{-.6cm}\includegraphics[scale=.18]{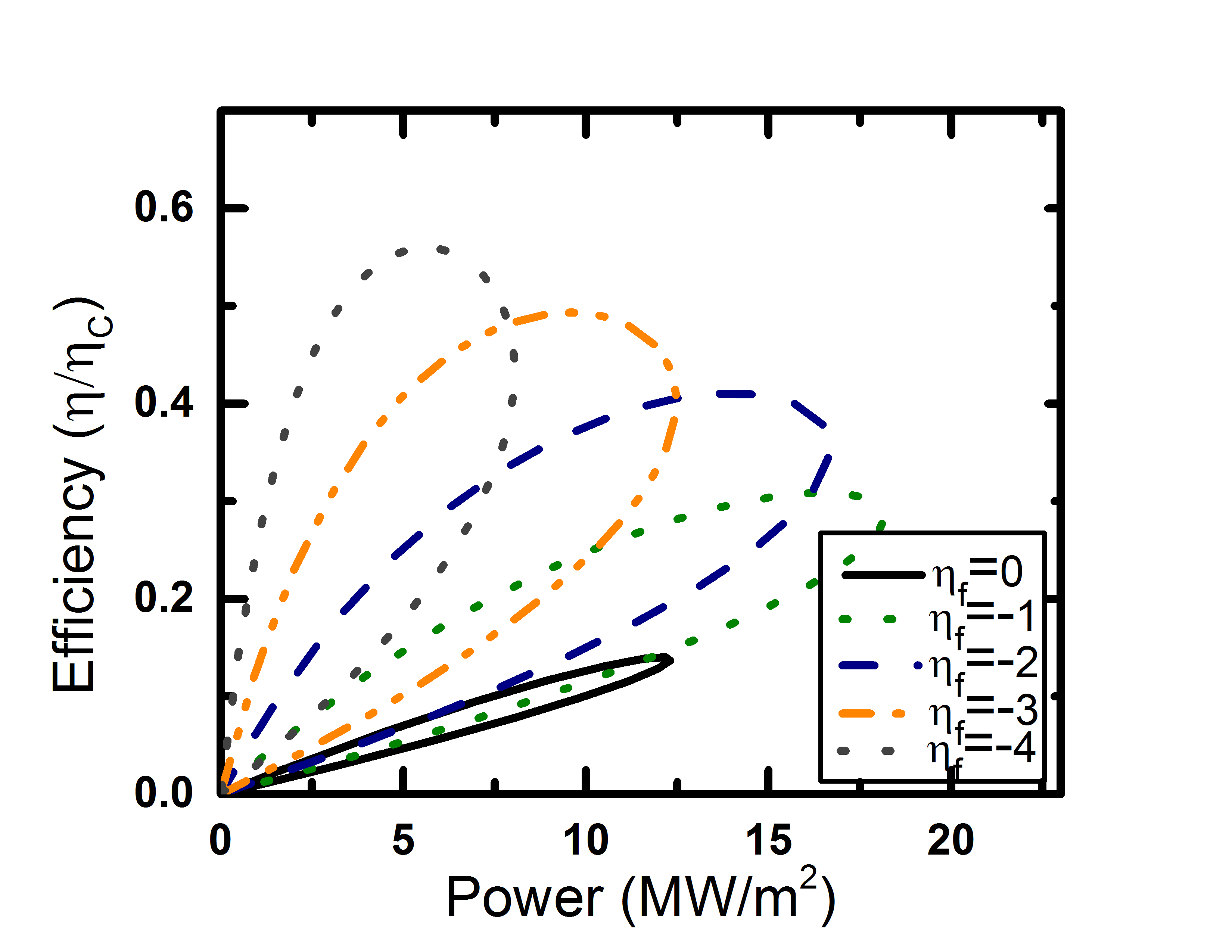}}
\subfigure[]{\hspace{-.5cm}\includegraphics[scale=.18]{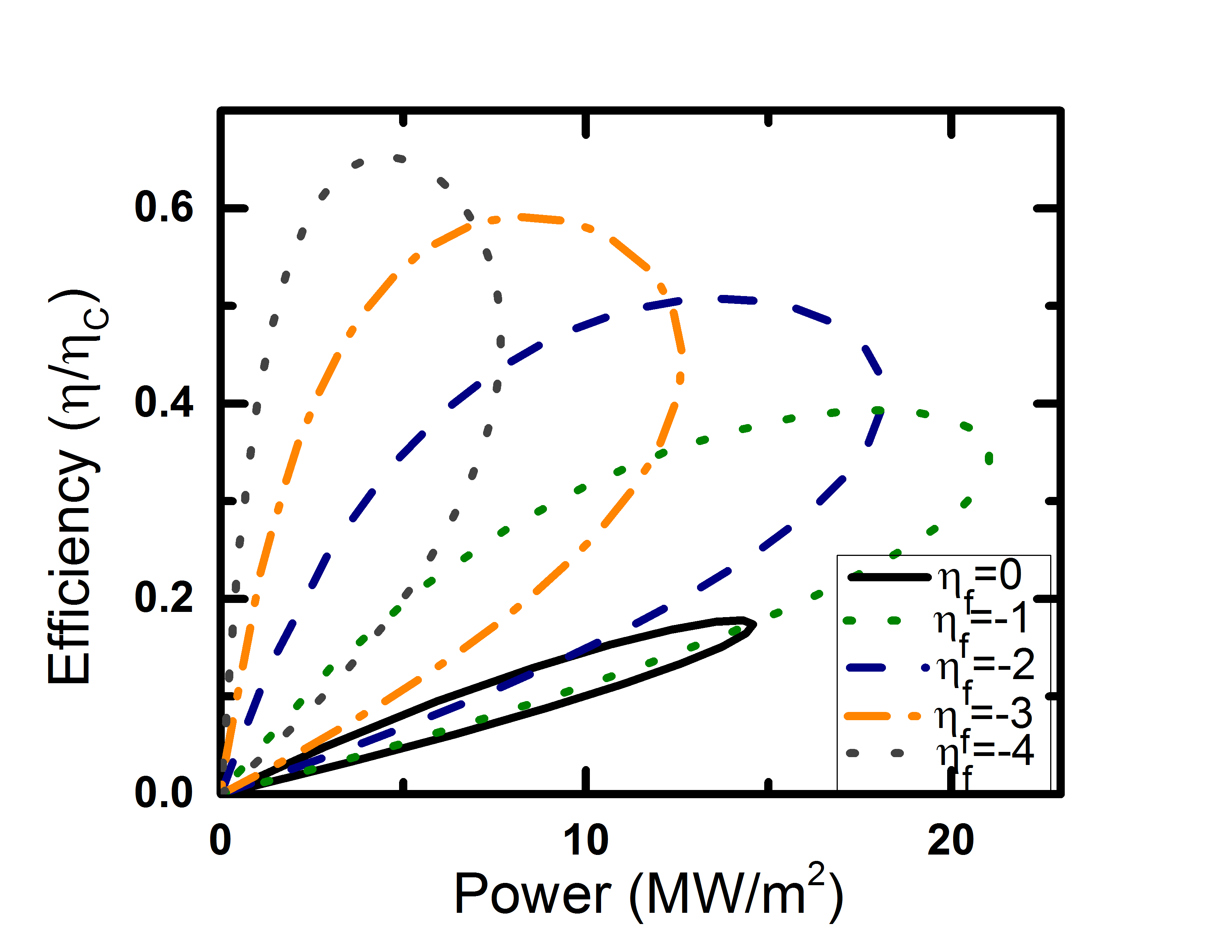}
}\subfigure[]{\hspace{-.5cm}\includegraphics[scale=.18]{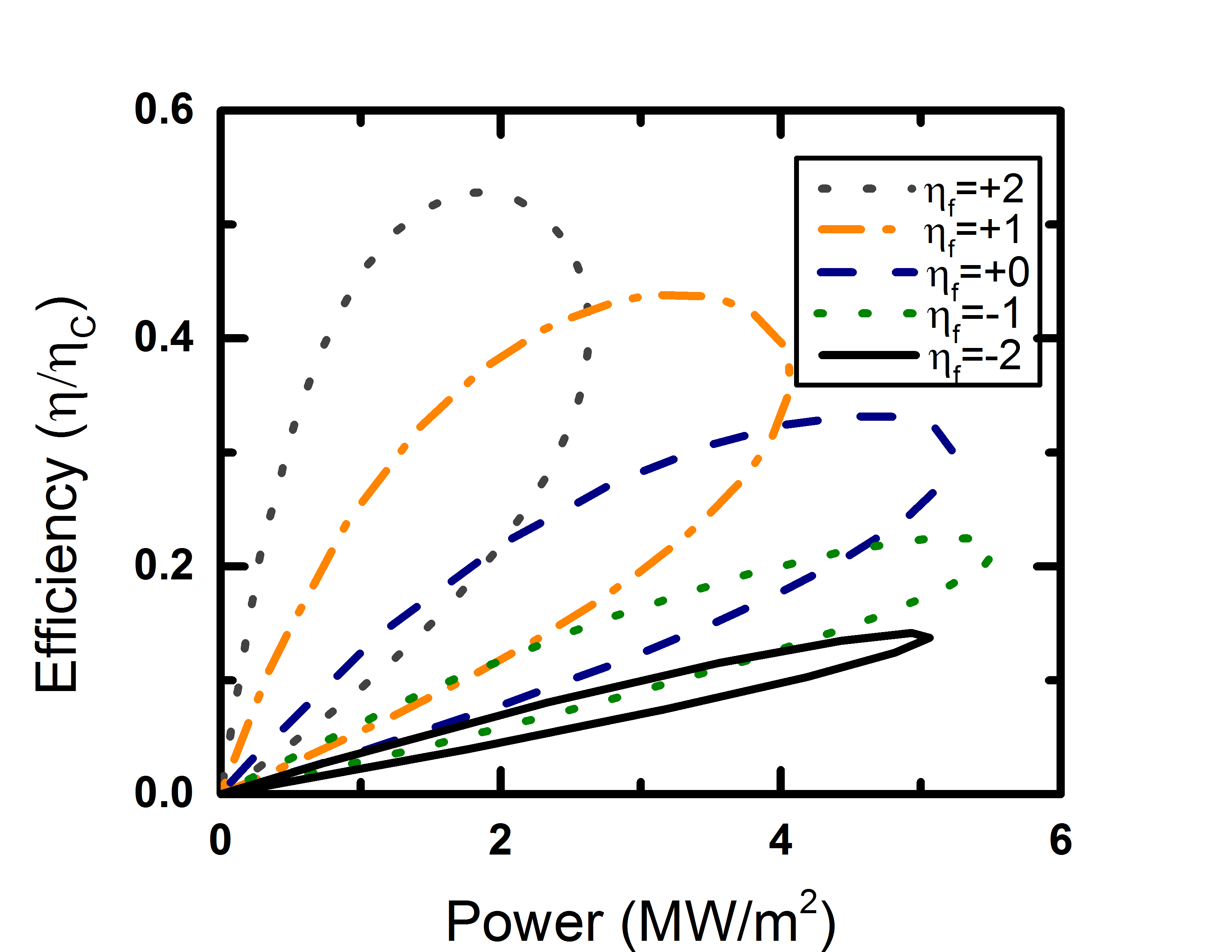}}
\subfigure[]{\hspace{-.6cm}\includegraphics[scale=.18]{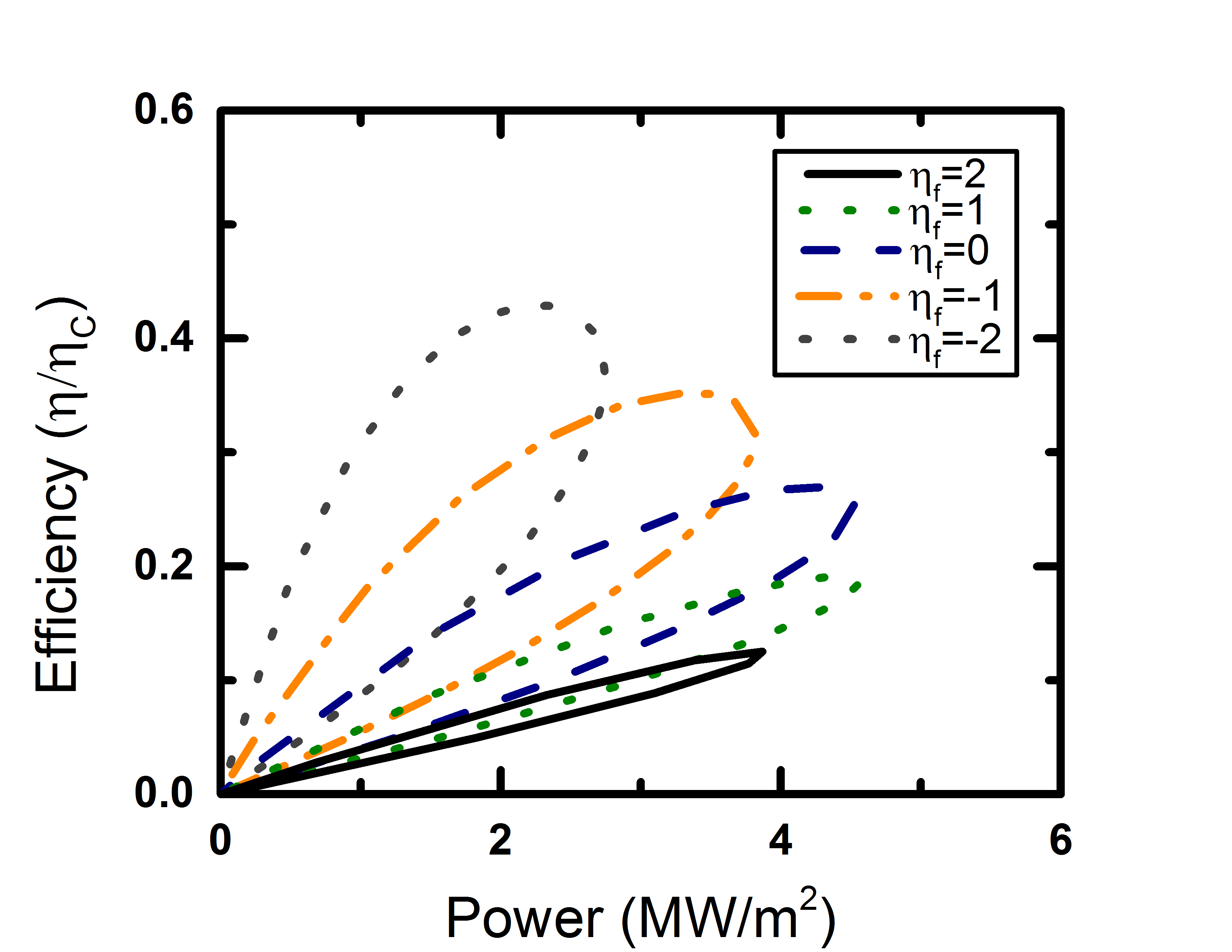}
}\subfigure[]{\hspace{-.5cm}\includegraphics[scale=.18]{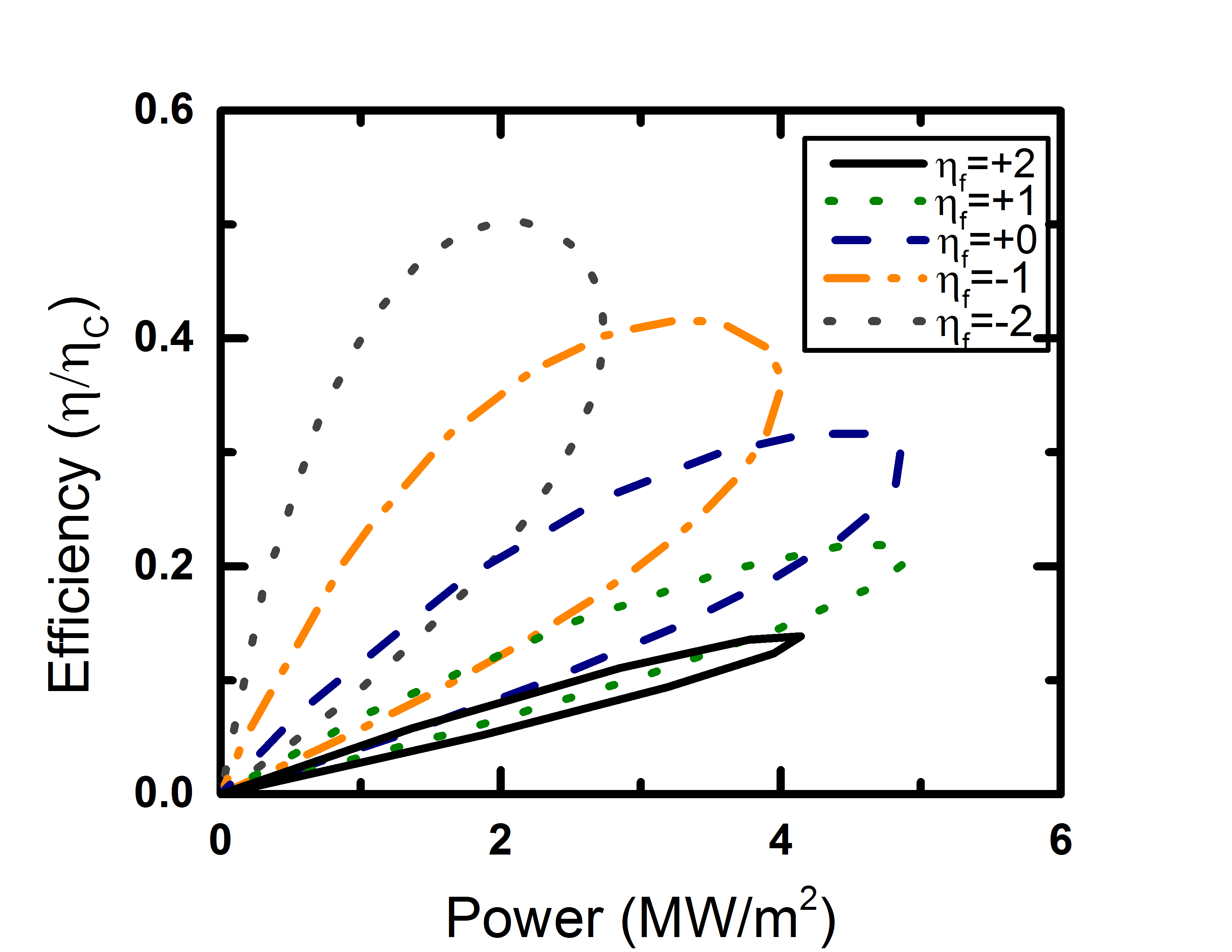}}
\subfigure[]{\hspace{-.6cm}\includegraphics[scale=.18]{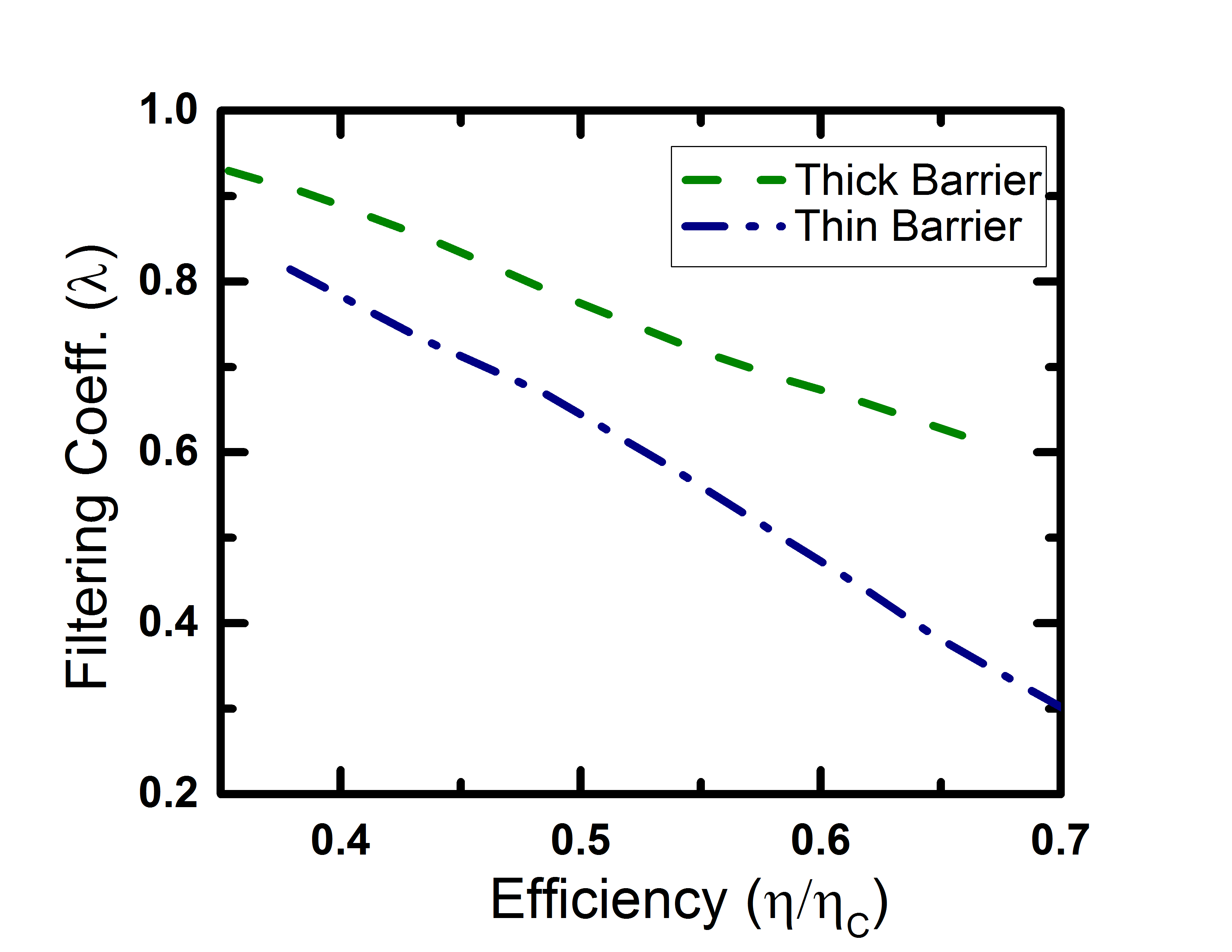}
}\subfigure[]{\hspace{-.6cm}\includegraphics[scale=.18]{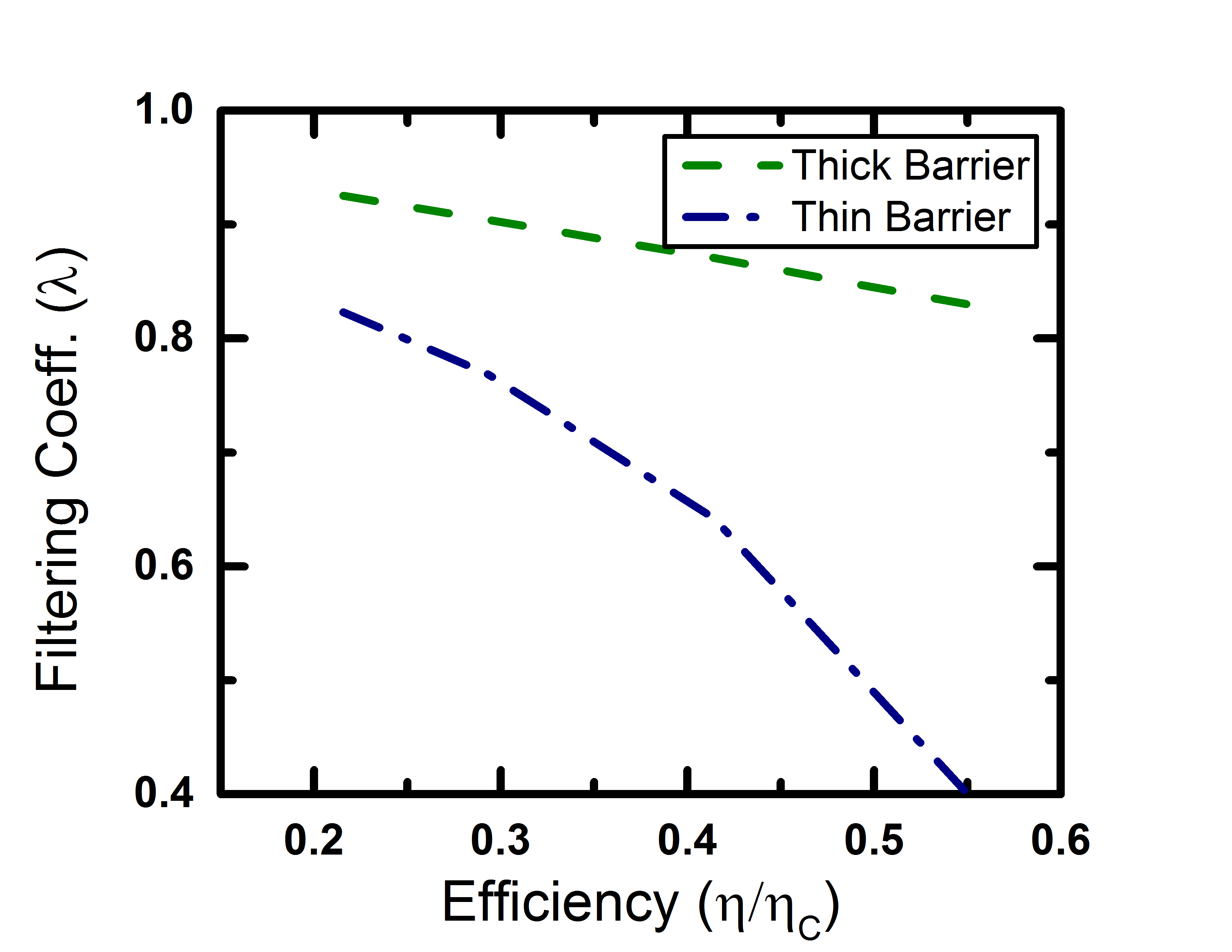}}
\caption{Power-efficiency trade-off and energy filtering analysis for ballistic devices. (a, b, c)- Plot of power density versus efficiency  for a $20nm$ long and  $2.85nm \times 2.85nm$ ballistic square  nanowire generator (a) without  energy filtering, (b) with energy filtering via a thin Gaussian energy  barrier ($E_b=150meV,~\sigma_w=1.35nm$), (c) with energy filtering via a thick Gaussian energy barrier ($E_b=150meV,~\sigma_w=2.7nm)$.  (d, e, f)- Plot of power density versus efficiency  for  a ballistic bulk thermoelectric generator (d) without  energy filtering, (e) with energy filtering via a thin Gaussian energy barrier ($E_b=150meV,~\sigma_w=1.35nm)$, (f) with energy filtering via a thick Gaussian energy barrier ($E_b=150meV,~\sigma_w=2.7nm)$.  (g) Plot of filtering coefficient ($\lambda$) versus efficiency ($\eta/\eta_C$) for the barriers used in (b) and (c). (h) Plot of filtering coefficient ($\lambda$) versus efficiency ($\eta/\eta_C$) for the energy barriers used in (e) and (f). }
\label{fig:ballistic_nanowire}
\end{figure}
In ballistic devices, electronic scattering is coherent in nature. In this case, although the average Seebeck coefficient increases due to filtering, the conductance $G$ decreases in a way to render energy filtering of electrons somewhat useless for power generation. Our findings are consistent with a recent work \cite{whitney} which showed that maximum thermoelectric power generation in ballistic devices occurs for a step like transmission function. Despite an increase in the density of states/modes with energy in bulk conductors, the conductivity in such devices do not increase due to filtering of high energy electrons via metallic nanoinclusions/energy barriers. The conductance in such devices is determined by the limited number of channels at the top of the barrier where the probability of electron transmission is minimum. \\
\indent We plot in Fig.~\ref{fig:ballistic_nanowire}, the power-efficiency trade-off curves for ballistic nano wires, (Figs. 2(a), (b) and (c)), and bulk devices (Figs. 2(d),(e) and (f)), with and without energy filtering. Each trade-off curve is plotted at a fixed value of the reduced Fermi energy, $\eta_f=\frac{E_c+E_b-\mu_0}{k_BT}$, where $E_c$ is the conduction band edge and $E_b$ is the height of the energy barrier. Efficiencies are evaluated with respect to the Carnot efficiency $\eta_C=1-T_C/T_H$. The generated power density for devices without energy filtering is found to be greater than that obtained with filtering as noted in the plotted trends of the filtering coefficient in  Figs. \ref{fig:ballistic_nanowire} (g) and (h), demonstrating the non-utility of energy filtering in such cases. An important point to note is that a thick barrier shows an improved filtering co-efficient compared to a thin barrier indicating the importance of a sharp energy cut-off for energy filtering.  The filtering coefficients of such devices decrease with increasing efficiency. This observation can be explained by the fact that at high efficiency the electrons near the Fermi level do not contribute towards conduction. Only a few electrons near the top of the barrier contribute to the conduction. With the transmission probability of such electrons being less than one, the filtering coefficient suffers drastically. We hence conclude that for ballistic devices, maximum power generation is achieved in cases where the transmission function is given by 
\begin{equation}
    T(\overrightarrow{k_{m}},E)= 
\begin{cases}
    1,& \text{if } E\geq E_m\\
    0,  & \text{if }  E<E_m,
\end{cases}
\end{equation}
where $E_m=\frac{\hslash^2k_m^2}{2m_t}$ is the minimum sub-band energy. In case of energy filtering with a barrier, the generated power would be the  same as the above case if 
\begin{equation}
    T(\overrightarrow{k_{m}},E)= 
\begin{cases}
    1,& \text{if } E\geq E_m+E_b\\
    0,  & \text{if }  E<E_m+E_b.
\end{cases}
\end{equation}

\begin{figure}[]
\subfigure[]{\hspace{-.5cm}\includegraphics[scale=.18]{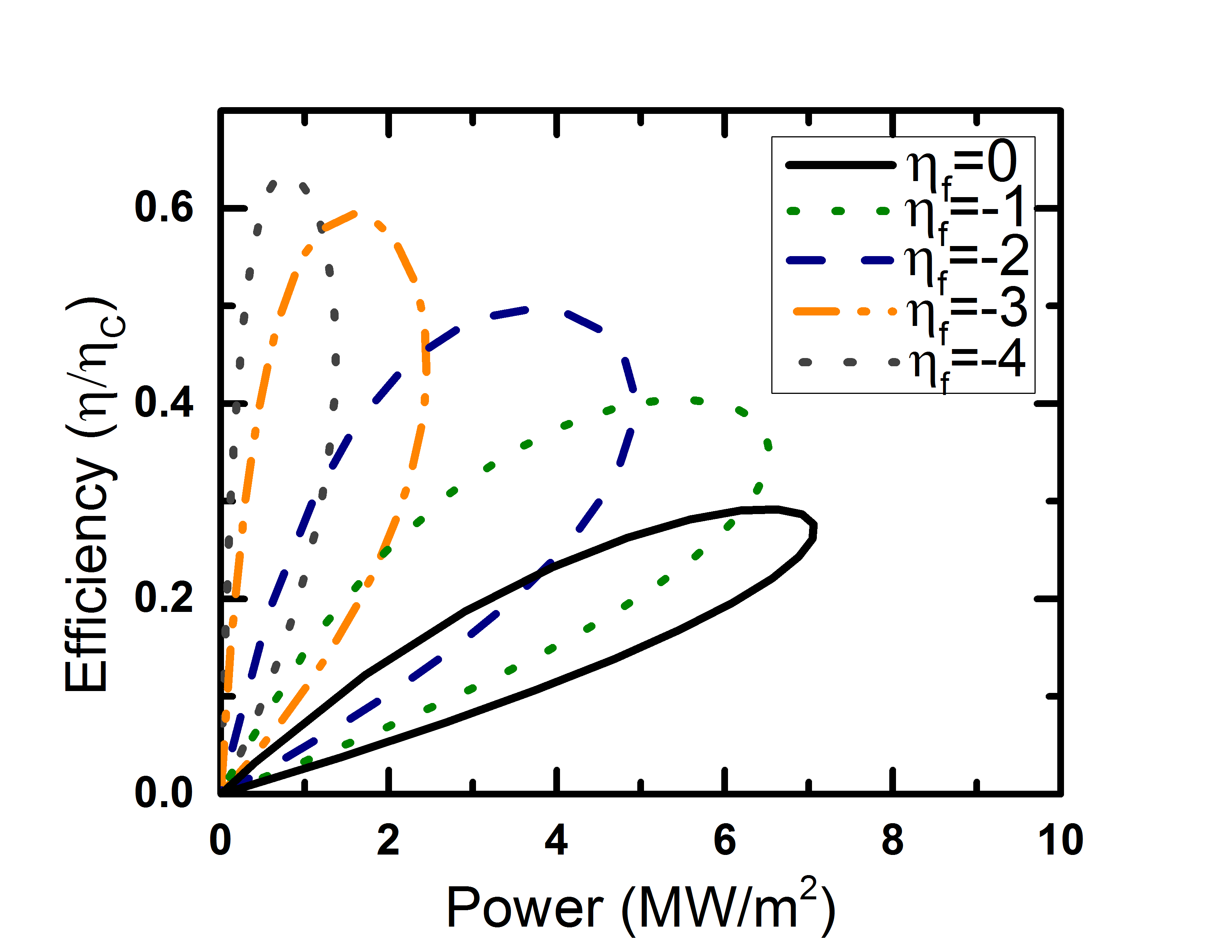}
}\subfigure[]{\hspace{-.6cm}\includegraphics[scale=.18]{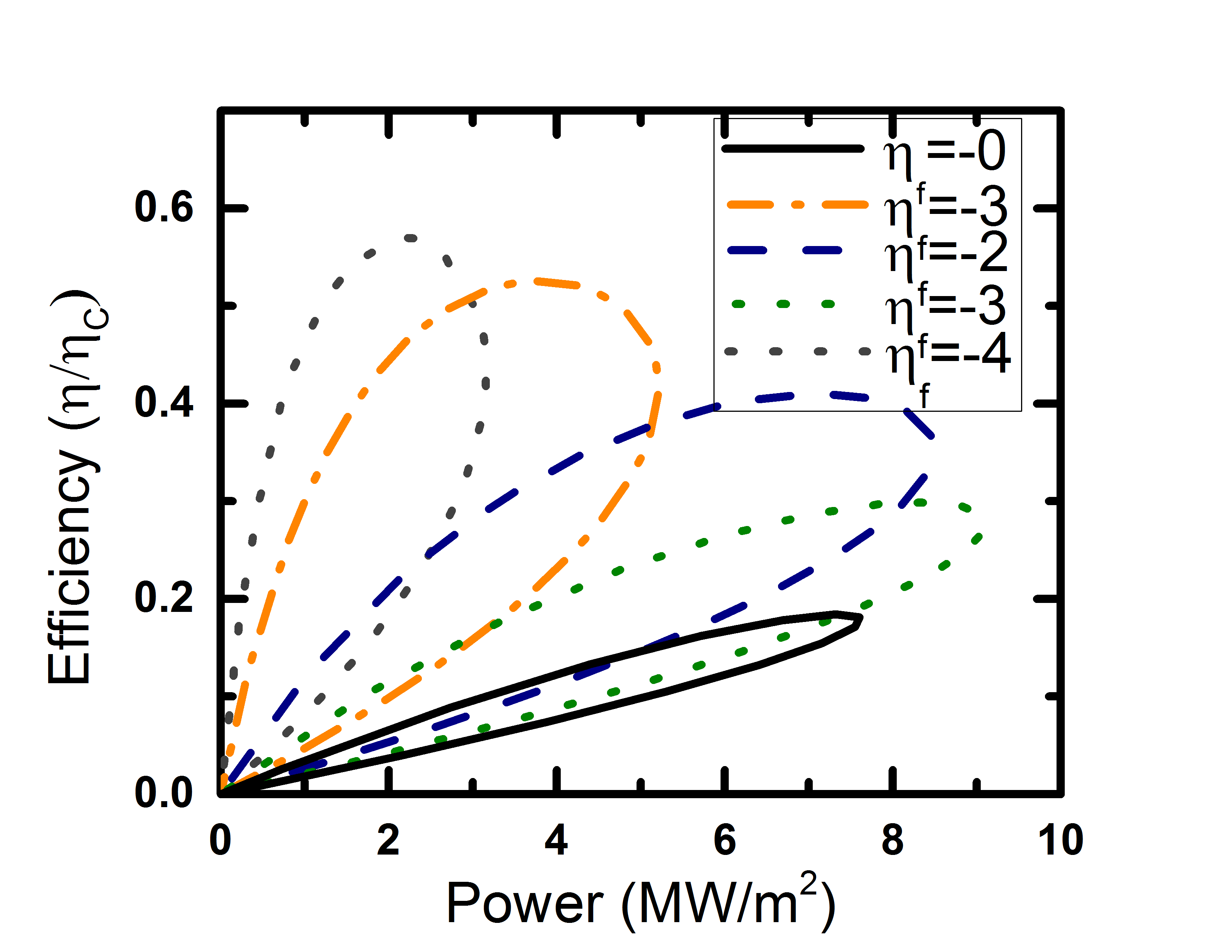}}
\subfigure[]{\hspace{-.5cm}\includegraphics[scale=.18]{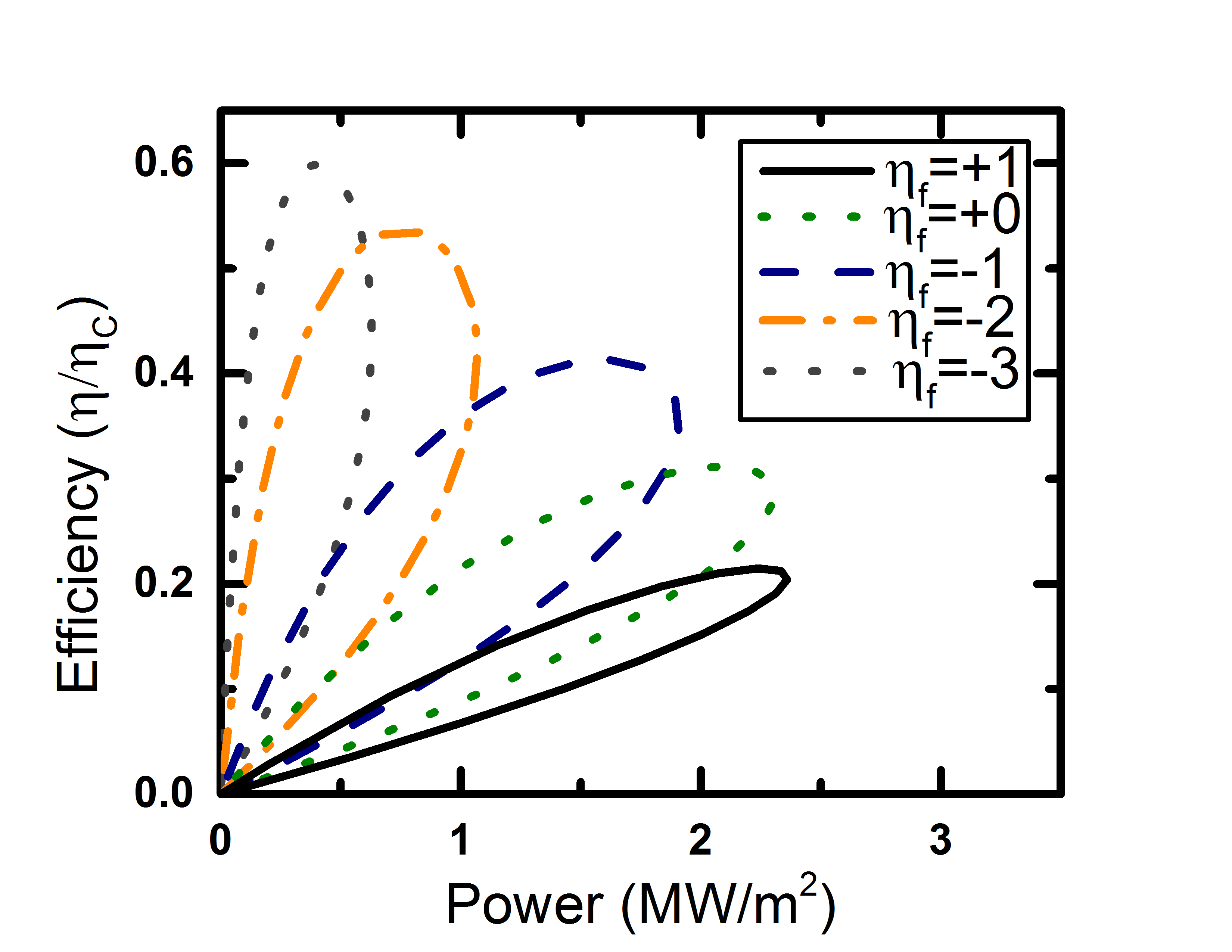}
}\subfigure[]{\hspace{-.6cm}\includegraphics[scale=.18]{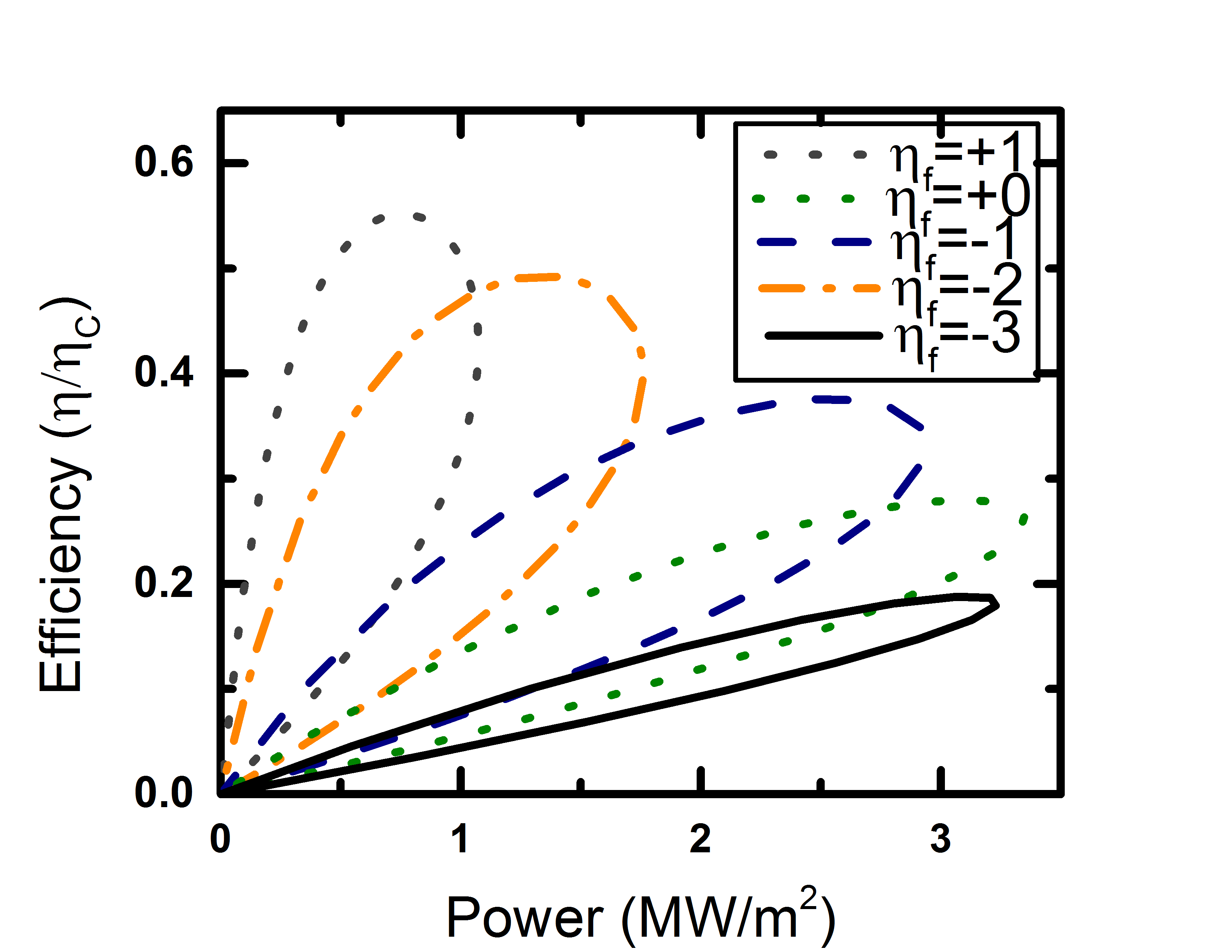}}
\subfigure[]{\hspace{-.5cm}\includegraphics[scale=.18]{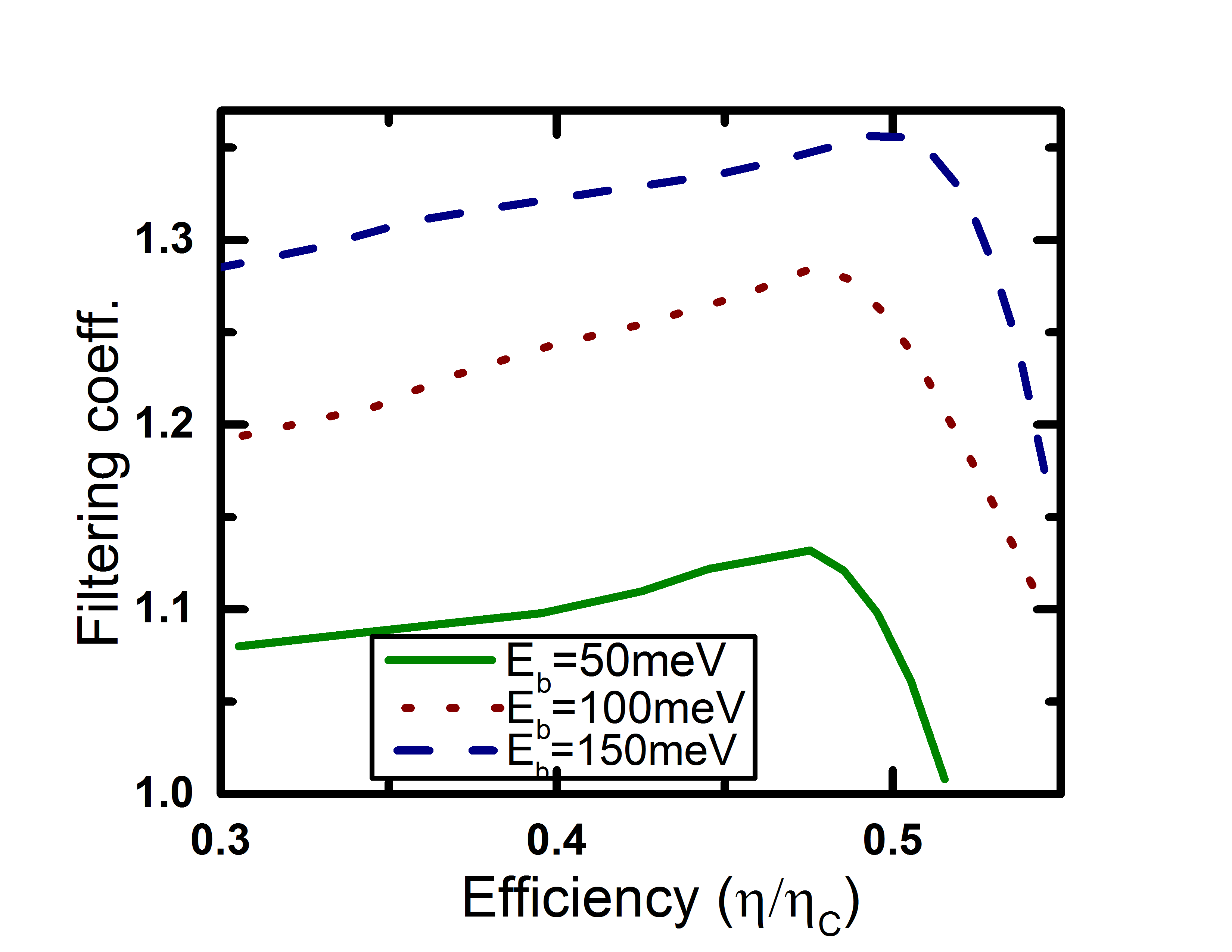}
}\subfigure[]{\hspace{-.6cm}\includegraphics[scale=.18]{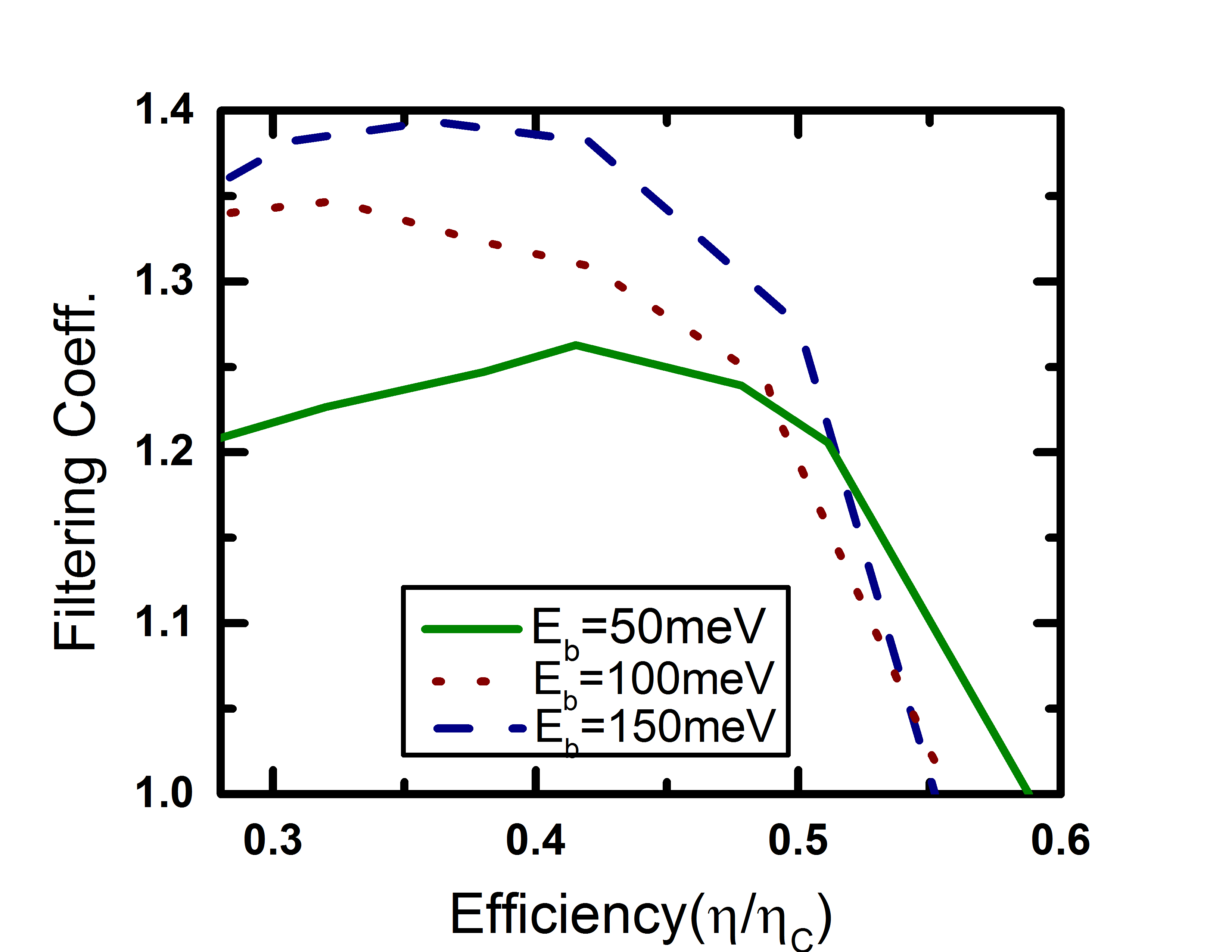}}
\caption{Power efficiency trade-off in the presence of incoherent acoustic phonon scattering. (a, b ,c, d)-Plots of power density versus efficiency curves at various values of $\eta_f$ for devices dominated by incoherent acoustic phonon scattering in case of (a) a single moded square nanowire of length $20nm$ and width $2.85nm$ without energy filtering, (b) a nanowire  of the same dimensions used in (a)  with energy filtering via a Gaussian energy barrier ($\sigma_w=2.7nm$, $E_b=150meV$), (c) bulk generator of length $20nm$ without energy filtering, and (d) bulk generator of the same dimensions used in (c) with energy filtering via a Gaussian energy barrier ($\sigma_w=2.7nm$, $E_b=150meV$). (e,f)-Plots of filtering coefficient $(\lambda) $ versus efficiency ($\eta/\eta_C$) in case of (e) a $2.85nm$ wide square nanowire of length $20nm$ embedded with a Gaussian energy barrier ($\sigma_w=2.7nm$),  (f) bulk generator of length $20nm$ embedded with a Gaussian energy barrier ($\sigma_w=2.7nm$).  }
\label{fig:nonballistic_nanowire}
\end{figure}

\subsection{Energy filtering with incoherent  scattering} \label{incoherent}
Incoherent scattering is characterized by a loss of overall momentum thereby contributing to the resistance of the device. In this section, we show that the enhancement of generated power due to energy filtering  is a characteristic of systems dominated by incoherent scattering. In Fig. \ref{fig:nonballistic_nanowire}, we plot the generated power density versus efficiency for nanowire (top panel) and bulk (bottom panel) thermoelectric generators with (right panel) and without (left panel) energy filtering with acoustic phonon scattering. An interesting point to note is that the maximum power density of nanowires without energy filtering occurs at $\eta_f=0$ instead of $\eta_f=-1$ for ballistic nanowires and nanowires with energy filtering. This occurs because electrons are almost immobile near $E=0$, due to the van-Hove singularity in conjunction with large scattering rates and low velocities. We plot in Fig. \ref{fig:nonballistic_nanowire} (e) and \ref{fig:nonballistic_nanowire} (f), the filtering coefficient  ($\lambda$) versus efficiency in case of  nanowire and bulk generators for various barrier heights.  The value of $\lambda$ increases with an increase in $E_b$. It can be shown that for single moded nanowires (Appendix \ref{appendix3}), this occurs due to an increase in the parameter $\Upsilon=v_z^2(E)\tau(E)D(E)$ with energy which results in the overall increase in current at a given voltage. An enhancement in generated power due to energy filtering is mainly dependent on two factors:- (i) increase in the number of electrons that has the potential to travel from source to drain contact per unit time, measured by the quantity $D(E)v_z(E)$ and (ii) decrease in the number of scattering  that each electrons suffers per unit length on average  while traveling from source to drain contact, measured by the parameter $\frac{1}{v_z(E)\tau(E)}$.    The parameter $\Upsilon=v_z^2(E)\tau(E)D(E)$ combines these two factors and is directly related to the conductivity of the device.
\begin{equation}
\sigma(E)=v_z^2(E)\tau(E)D(E)\Big\{-\frac{\partial f}{\partial E}\Big\}
\label{eq:conductivity}
\end{equation} 
where $v_z(E)$, $\tau(E)$ and $D(E)$ are the velocity of electrons in the transport direction, relaxation time and density of states at energy $E$ respectively.  % The slight increase in $\lambda$ due with increase in efficiency can be explained by the fact that at high efficiency, the fermi level in a nano wire lies a few $k_BT$ below the conduction band edge. The electrons contributing to conduction are available near the bottom of the conduction band where mobility is very low resulting in a sharp decrease in generated power with increase of efficiency for devices without energy filtering.%
With increase in efficiency beyond a certain point, the filtering coefficient sharply decreases due to the effect of imperfect filtering, particularly due to a smooth cutoff energy and partial suppression of transmission probability,  in the presence of a Gaussian energy barrier. \\
\indent For bulk thermoelectric generators, it can be shown that for acoustic phonon scattering, the parameter $\Upsilon=v_z^2(E)\tau(E)D_{con}(E)$ increases with  energy due to increase in velocity as well as increase in intermode current due to intermode coupling (Fig. \ref{fig:intermode2}), where $D_{con}(E)$ is the density of states contributing to the current flow.

\subsection{Energy filtering due to higher order scattering mechanisms}\label{order}
For acoustic phonon scattering, $\tau(E)=k_oE^{\frac{1}{2}}$ for nanowires and $\tau(E)=k_oE^{-\frac{1}{2}}$ for bulk devices. In section \ref{incoherent}, we have already shown that energy filtering in devices dominated by incoherent acoustic phonon scattering leads to an enhancement in the generated power. However, in practical devices, higher order scattering may exist such that $\tau(E)=k_oE^r$ with $r<{\frac{1}{2}}$ for nanowires and $r<{-\frac{1}{2}}$ for bulk generators respectively. Then, the question naturally arises, is there a minimum value of $r$   for which energy filtering enhances power generation? \\
\indent It can be shown theoretically  that in case of perfect energy filtering (sharp cut-off energy) for single-moded nanowires, filtering coefficient is always enhanced with energy filtering  when the parameter $\Upsilon=v^2(E)\tau(E)D(E)$ is an increasing function of energy (Appendix \ref{appendix3}).
 For a 1-D nanowire, $v_z^2(E)=\frac{2E_z}{m_l}$ and $D(E_z)=\frac{1}{\hslash\pi}\sqrt{\frac{m_l}{2E_z}}$. Assuming $\tau$ to be of the form $\tau(E_z)= kE_z^{r}$, for perfect filtering with sharp cut-off energy, power generation is enhanced for $r \geq r_{min}$ where $r_{min}=-\frac{1}{2}$. For imperfect filtering  $r_{min} > -\frac{1}{2}$  (See Appendix \ref{appendix3}). In such cases, $r_{min}$ is a function of efficiency of operation.   An analytical calculation of $r_{min}$ for bulk generators is not so trivial due to intermode coupling. For energy filtering in  bulk generators dominated by moderate electron-phonon scattering, the density of states contributing to  the conductivity is ill defined due to partial momentum conservation. However, we can draw some conclusions on the upper bounds of $r_{min}$ under the assumption of independent modes. In case of perfect lateral momentum conservation, i.e., independent modes, it can be shown that for perfect filtering with $E_b=150meV$,  $r_{min} \approx -0.6$ (Sec \ref{appendix3}) \\
 \begin{figure}[]
\subfigure[]{\hspace{-.5cm}\includegraphics[scale=.18]{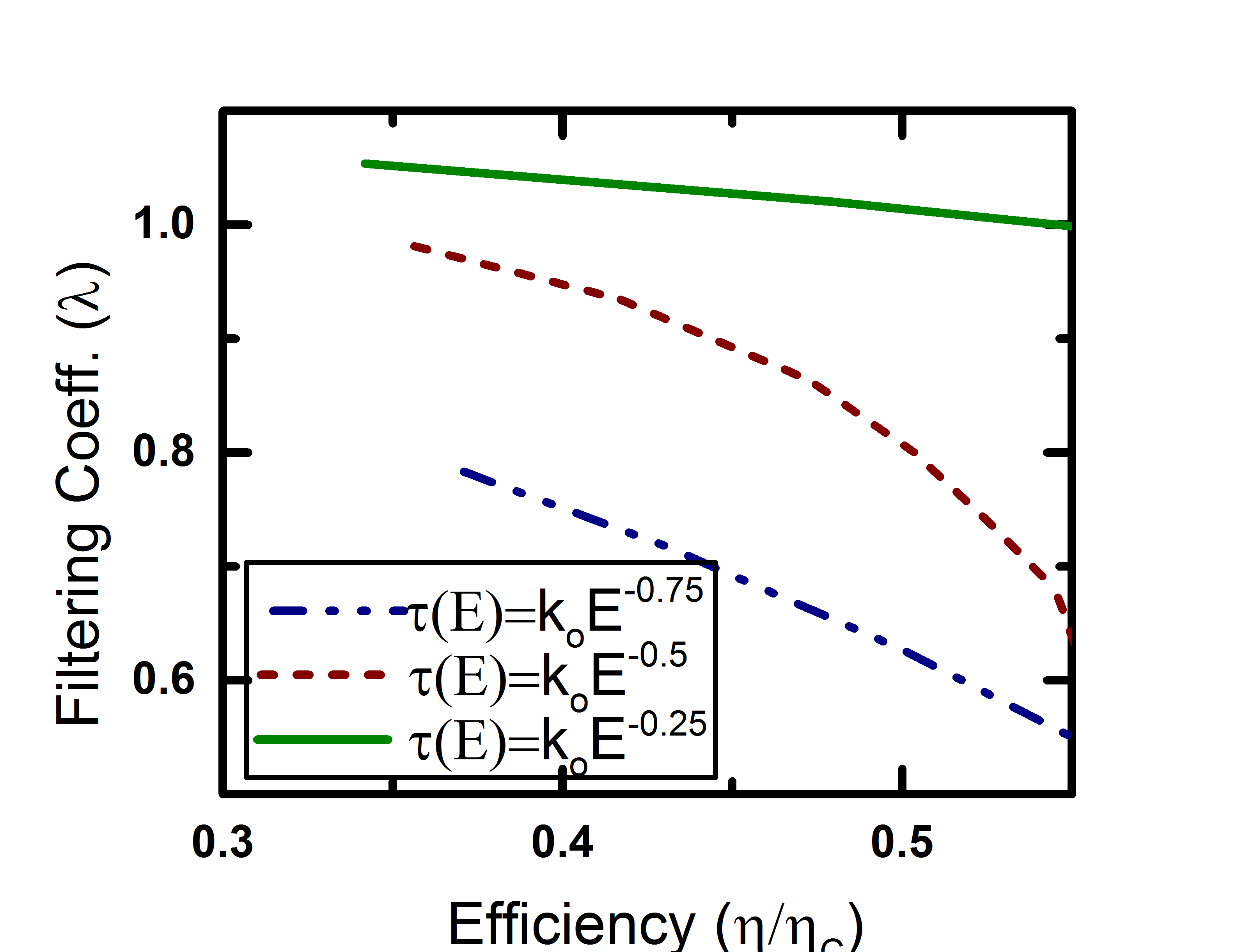}
}\subfigure[]{\hspace{-.6cm}\includegraphics[scale=.18]{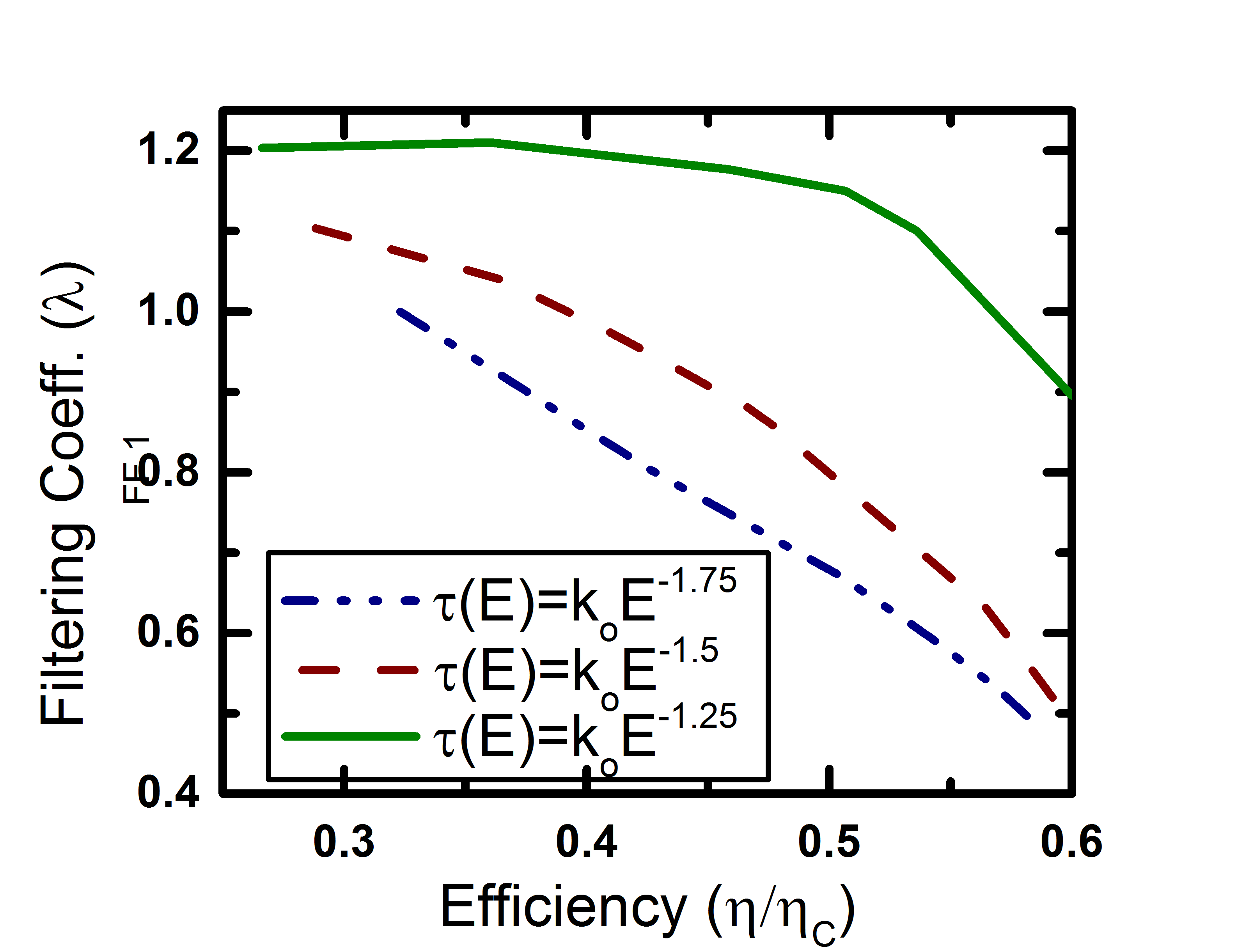}}
\caption{Plot of filtering coefficient ($\lambda$) versus efficiency ($\eta/\eta_C$) with higher order scattering ($\tau(E)=k_oE^r$). Plots are shown for  (a)  single moded square nanowires of width $2.85nm$ and length $20nm$ with $r=-0.75$, $r=-0.5$ and $r=-0.25$ and   (b)  bulk thermoelectric generators of length $20nm$ with $r=-1.75$, $r=-1.5$ and $r=-1.25$. The  decrease in filtering coefficient of nanowires at $r=-0.5$ compared to theoretical predictions ($\lambda_{ideal}=1$)  is due to imperfect filtering (partial transmission of electrons). }
\label{fig:other_pros}
\end{figure}
\indent The filtering coefficient ($\lambda$) versus efficiency ($\eta/\eta_C$) plots for nanowire and bulk generators with higher order scattering are shown in Fig. \ref{fig:other_pros}. 
It is demonstrated in  Fig. \ref{fig:other_pros} (b)  that for bulk thermoelectric generators, even with imperfect energy filtering, power generation is enhanced for $r<-0.6$. To explain this, we need to delve into the details of intermode scattering and understand its contribution to power generation.
\subsection{Role of intermode coupling in enhancing power generation for bulk generators}\label{intermode}

\begin{figure}[!htb]
\subfigure{\fbox{\includegraphics[scale=.245]{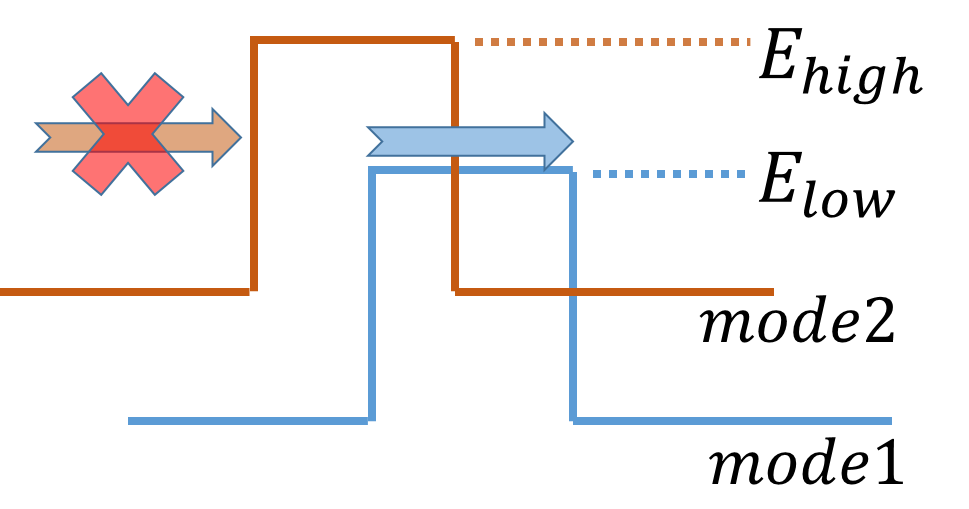}}
}\subfigure{\fbox{\includegraphics[scale=.245]{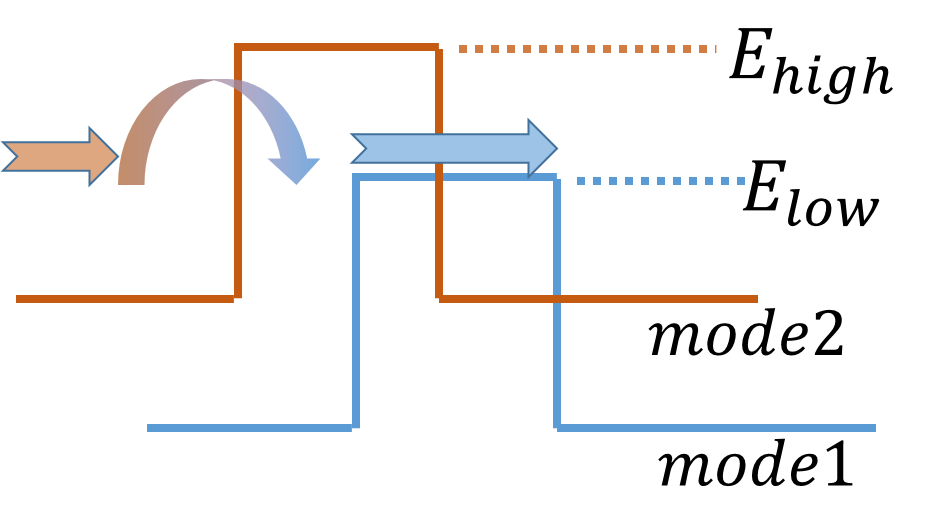}}}
\caption{Schematic diagram illustrating electronic transport through a device with two modes (a) in the absence of electron-phonon interaction and (b) in the presence of electron-phonon interaction.  In the absence of electron-phonon interaction, the electrons from the higher energy mode are completely reflected from the barrier while the electrons with the same energy from the lower energy mode are transmitted.  The situation changes in the presence of electron-phonon interaction, when the electrons from the higher energy  mode are kicked  to the lower energy mode via transfer of momentum to phonons. Such processes contribute significantly to the generated current and hence power.}
\label{fig:intermode}
\end{figure}
In bulk thermoelectric generators, intermode scattering of electrons have a finite contribution to the enhancement of generated power. In case of energy filtering in ballistic semiconductors, lateral momentum is conserved and electrons from the higher energy modes cannot contribute to power generation due to coherent  reflection from the barrier. However, in systems dominated by electron-phonon interaction, incoherent scattering can drive a finite current from the higher energy modes to the lower energy modes while breaking the  conservation of lateral momentum.   The concept of quasi equilibrium is a characteristic feature of diffusive systems and is  mediated by electron-phonon interaction. In case of energy filtering, the lower energy modes are driven out of equilibrium due to the current flow. In the absence of electron-phonon interaction, the higher energy modes are in equilibrium because no current can flow through the barrier. In the presence of electron-phonon interaction, electrons from the higher energy modes can flow to the lower energy modes to restore equilibrium and can contribute significantly to power generation. Such flow of intermode current from the high energy modes to the low energy modes occurs in the region between the source contact and the barrier interface. Hence, the intermode current is maximum if the length of the device between the source contact and barrier interface is longer than a few momentum relaxation lengths.
  \begin{figure}
 \subfigure{\includegraphics[scale=.3]{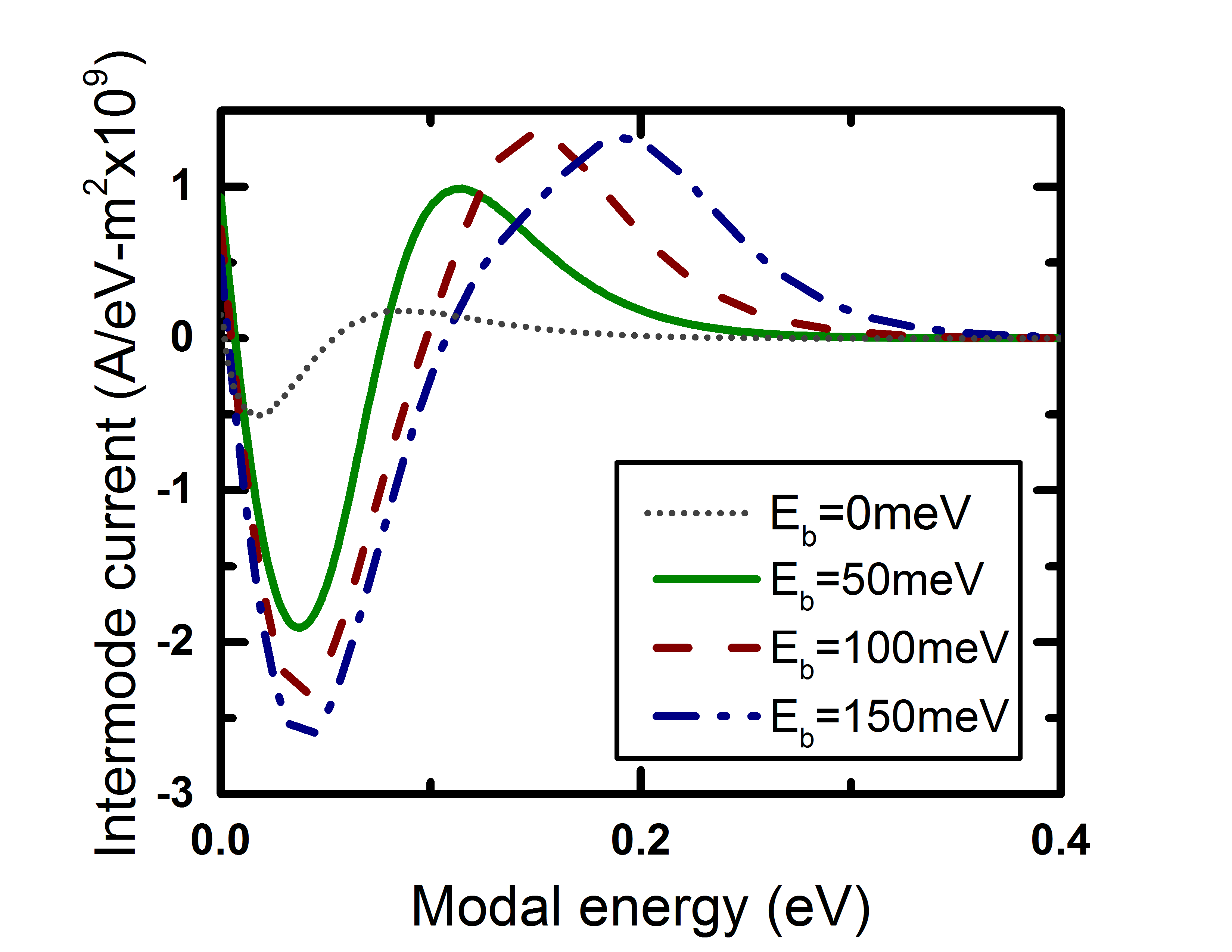}}

 \caption{Plot of intermode current  per unit area per unit energy  in case of  bulk generators at the maximum power. An increase in the   total intermode current flow with the height of the energy  barrier  ($E_b$)  contributes to the enhancement in power generation. Simulations are  carried out for a $20nm$ long device   embedded with a Gaussian energy barrier ($\sigma_w=2.7nm$)  taking acoustic phonon scattering into account. Negative value of intermode current corresponds to inflow of electrons while a positive value corresponds to outflow of electrons. }
 \label{fig:intermode1}
 \end{figure}
 We schematically illustrate in Fig. \ref{fig:intermode}  the simple case of electronic transport through a device in the presence of  two modes  with and without electron-phonon interaction. In Fig \ref{fig:intermode}.(a), the current in the energy range $E_{low}$ to $E_{high}$ is carried by $mode1$ while that in the energy range above $E_{high}$ is carried by both $mode1$ and $mode2$. However in Fig \ref{fig:intermode}.(b) a significant portion of the current in the energy range $E_{low}$ to $E_{high}$ is contributed by $mode2$ in addition to the  obvious contribution by $mode1$. Such contribution to the  current flow by $mode2$ results from electron-phonon interactions.  %In the absence of electron-phonon interaction, the modes $s1$ and $s2$ carries current independently. The electrons in the energy range between $E1$ and $E2$ in mode 2 does not contribute to the current flow, however, the electron population between $E1$ and $E2$ in $s1$ is driven out of equilibrium due to current flow. 
 The electron-phonon interaction, when switched on, drives the electrons in the energy range $E_{low}$ to $E_{high}$ from $mode2$ to $mode1$ to restore equilibrium, thus yielding a finite contribution to the current flow. For bulk thermoelectric generators the density of modes $\Big(M(E)=\frac{m^*}{2*\pi*\hslash^2}(E-E_C)\Big)$ and the rate of acoustic phonon scattering increase with  energy. The  increase in the density of modes with energy  and hence intermode electron flow contributes to the enhancement of generated power.

 \begin{figure}
  \subfigure{\includegraphics[scale=.3]{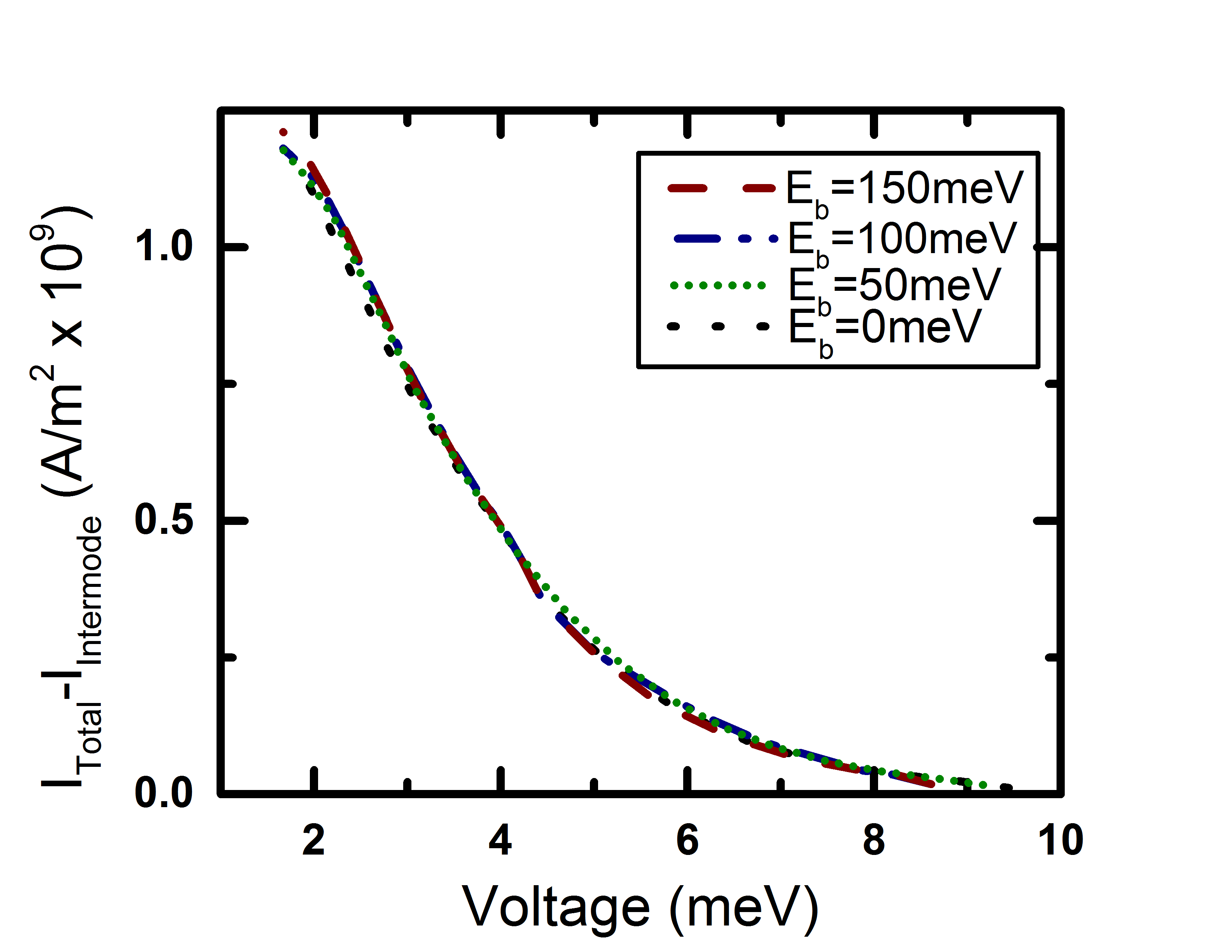}
  }
  
   \subfigure{\includegraphics[scale=.3]{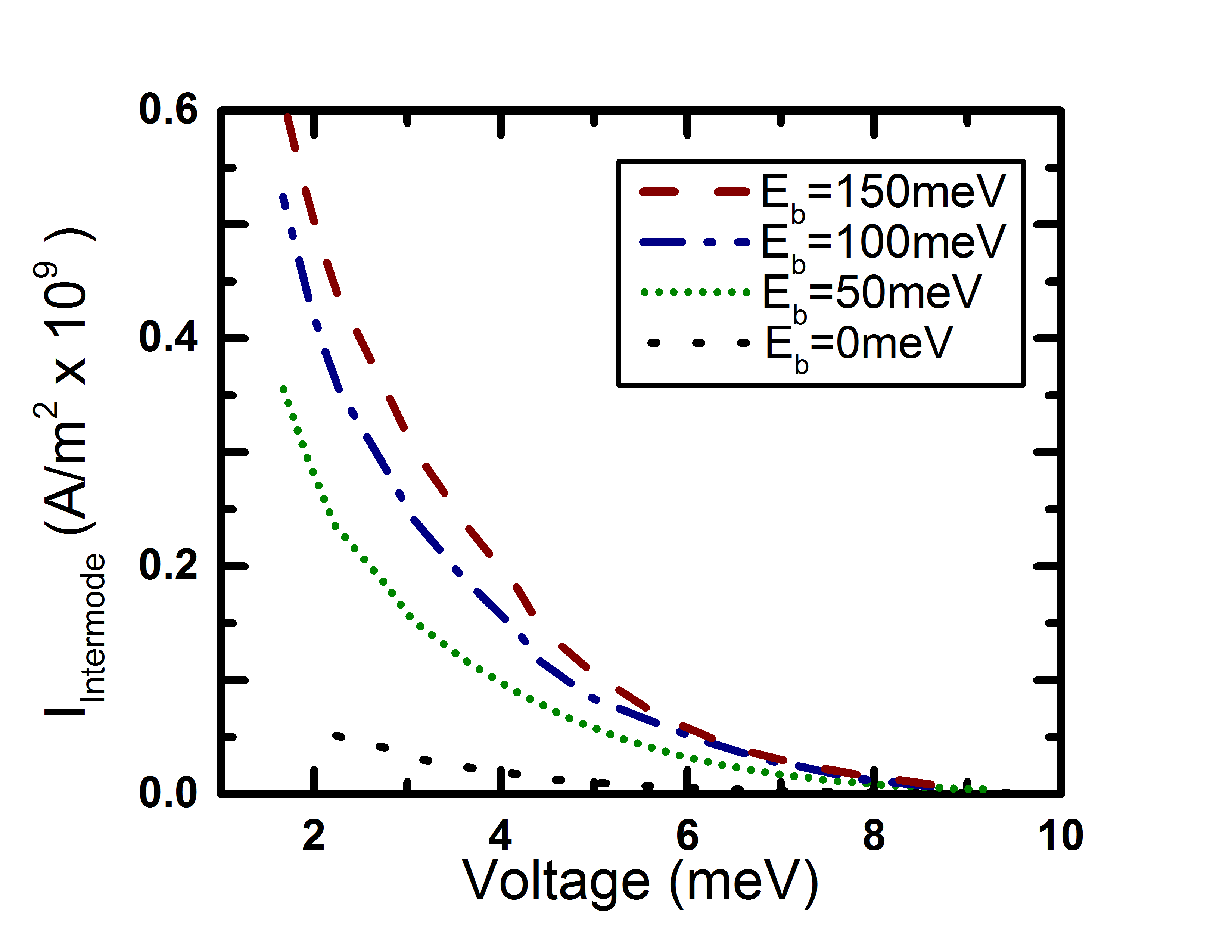}}

\caption{Plot of  (a) propagating  current  without the intermode current at the maximum power at a given voltage and (b) intermode current per unit area at the maximum power at a given voltage. Simulations are done for incoherent electronic transport  taking acoustic phonon scattering into account. The increase in intermode current at the maximum power at a given voltage  with barrier height ($E_b$) is the main factor behind the increase in generated power due to energy filtering.}
\label{fig:intermode2}
 \end{figure}

 \begin{figure}[]
\subfigure[]{\hspace{-.5cm}\includegraphics[scale=.18]{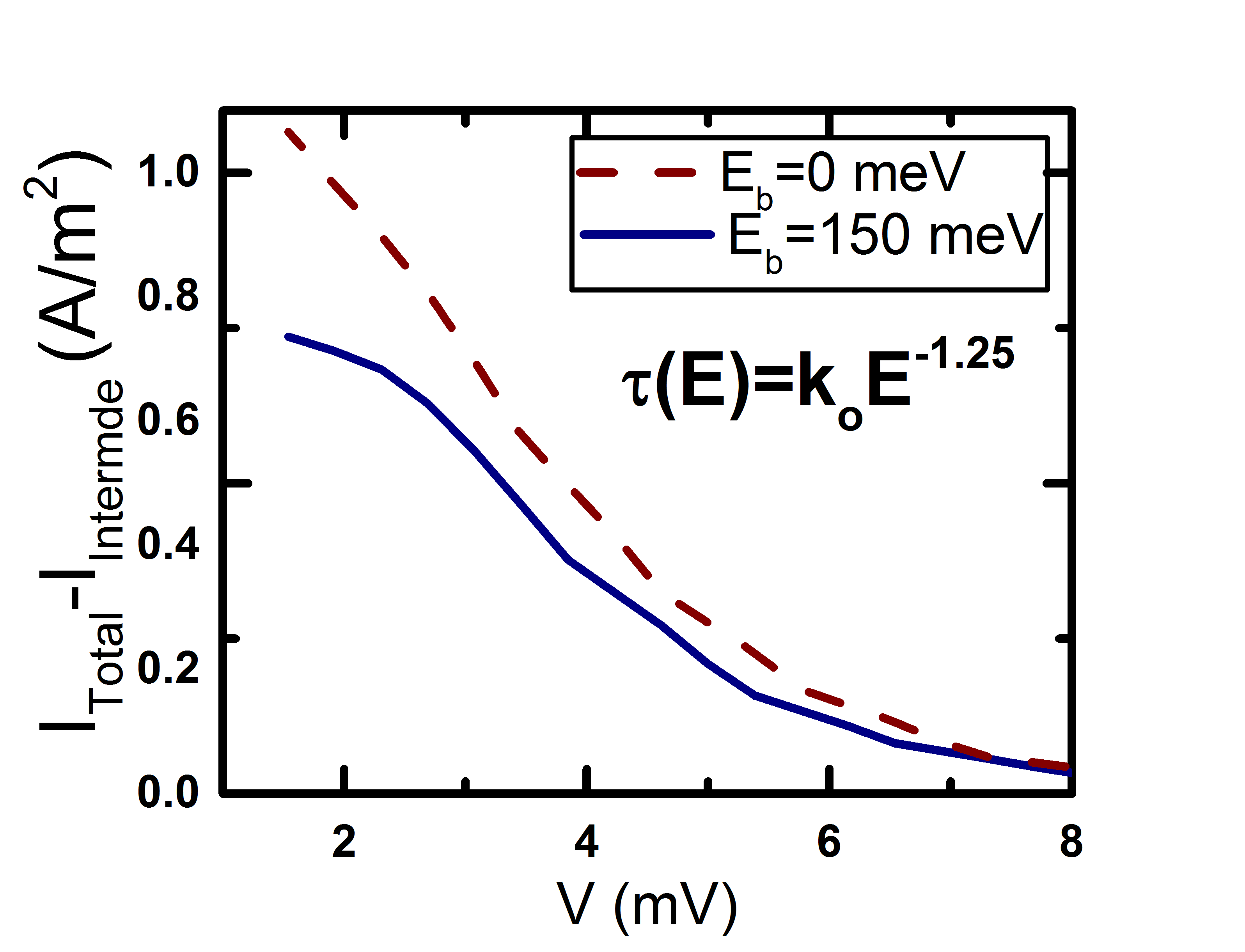}
}\subfigure[]{\hspace{-.6cm}\includegraphics[scale=.18]{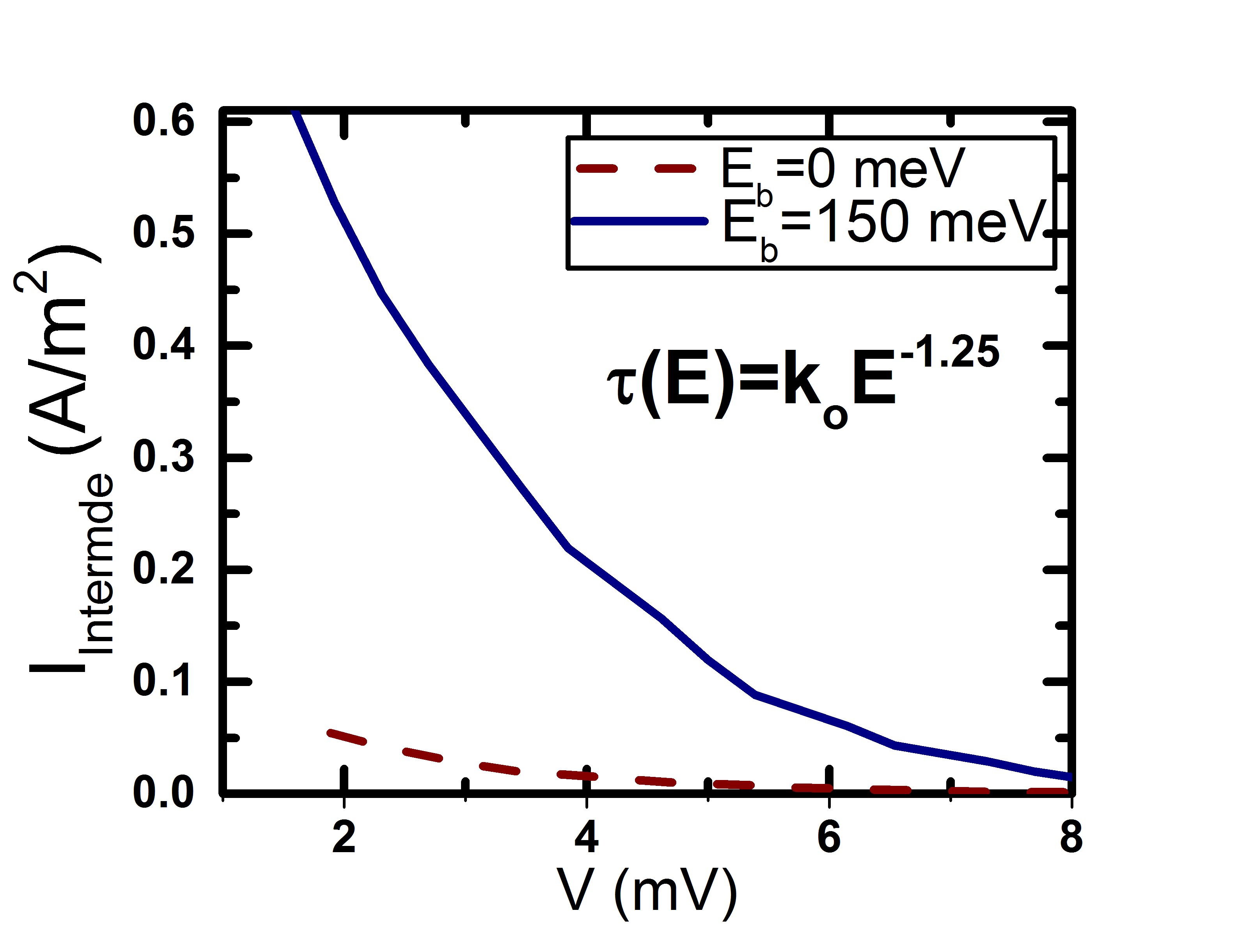}}
\subfigure[]{\hspace{-.5cm}\includegraphics[scale=.18]{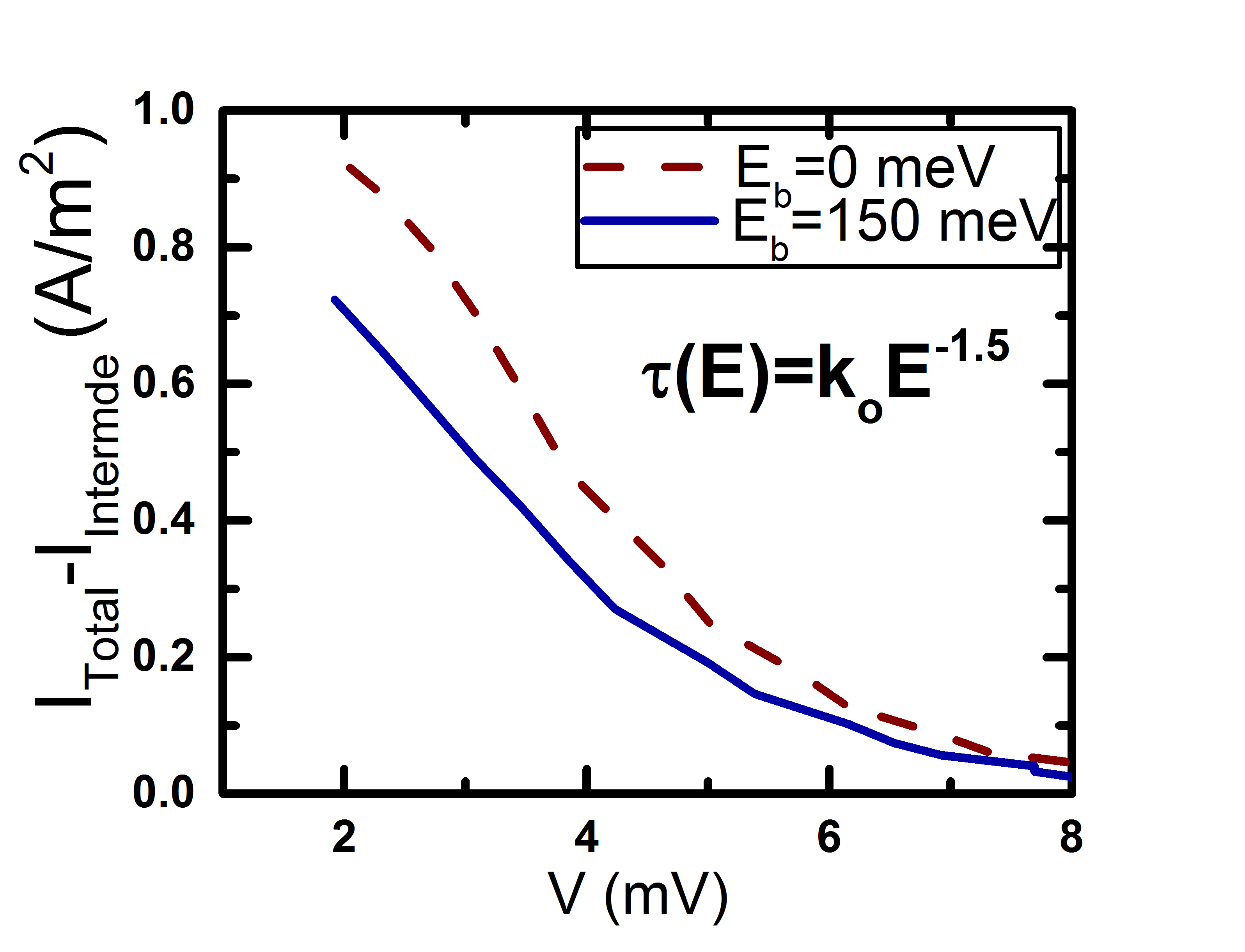}
}\subfigure[]{\hspace{-.6cm}\includegraphics[scale=.18]{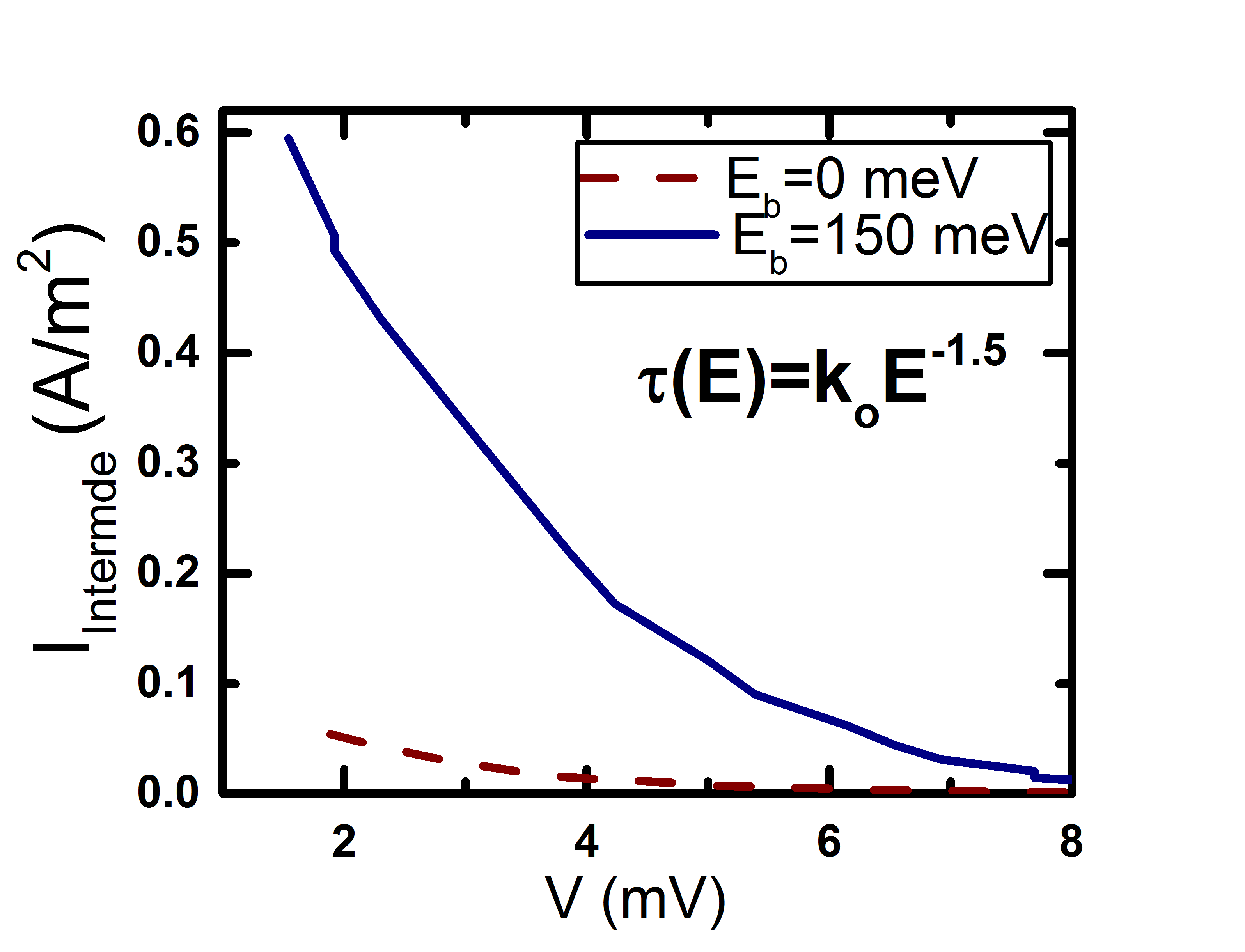}}
\subfigure[]{\hspace{-.5cm}\includegraphics[scale=.18]{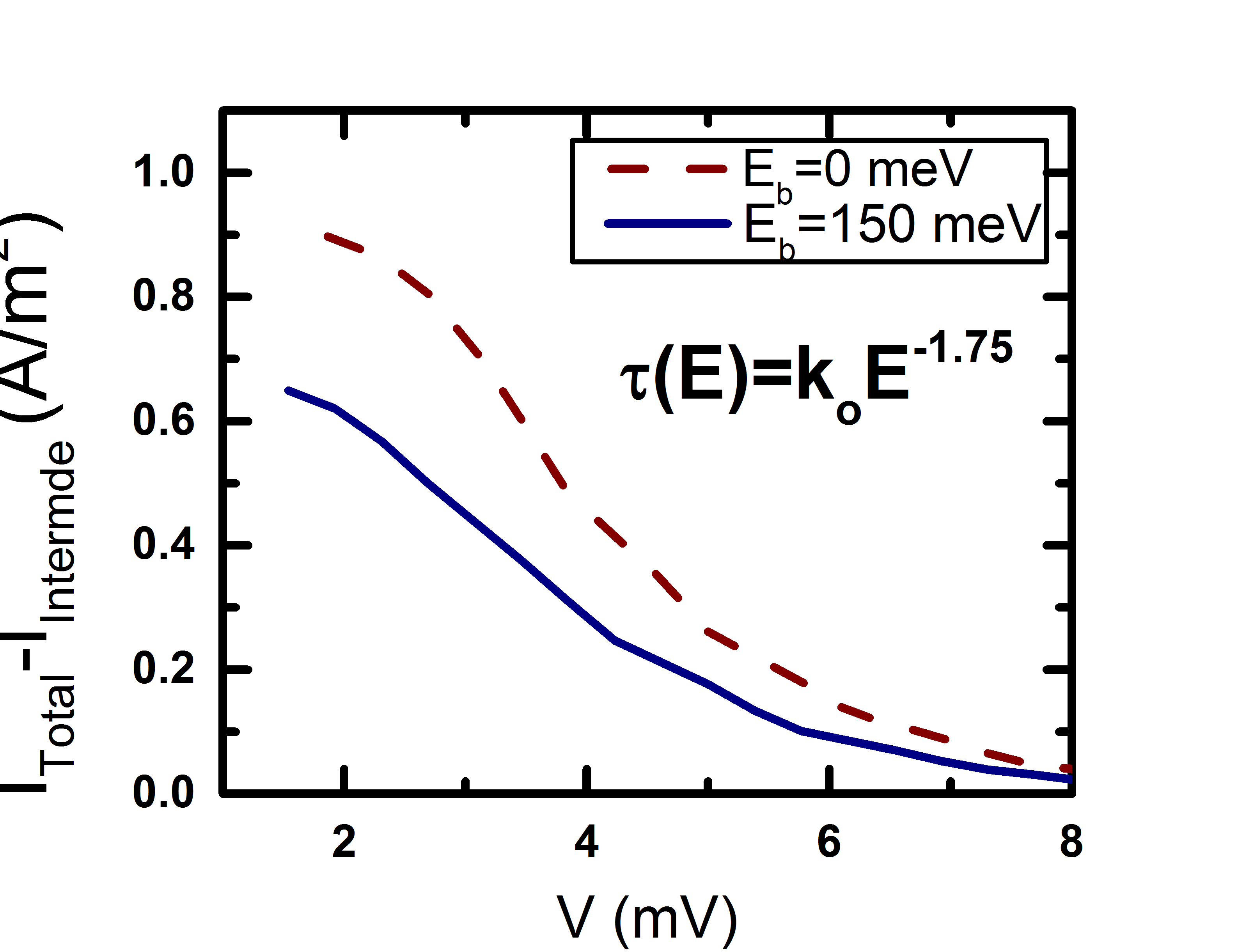}
}\subfigure[]{\hspace{-.6cm}\includegraphics[scale=.18]{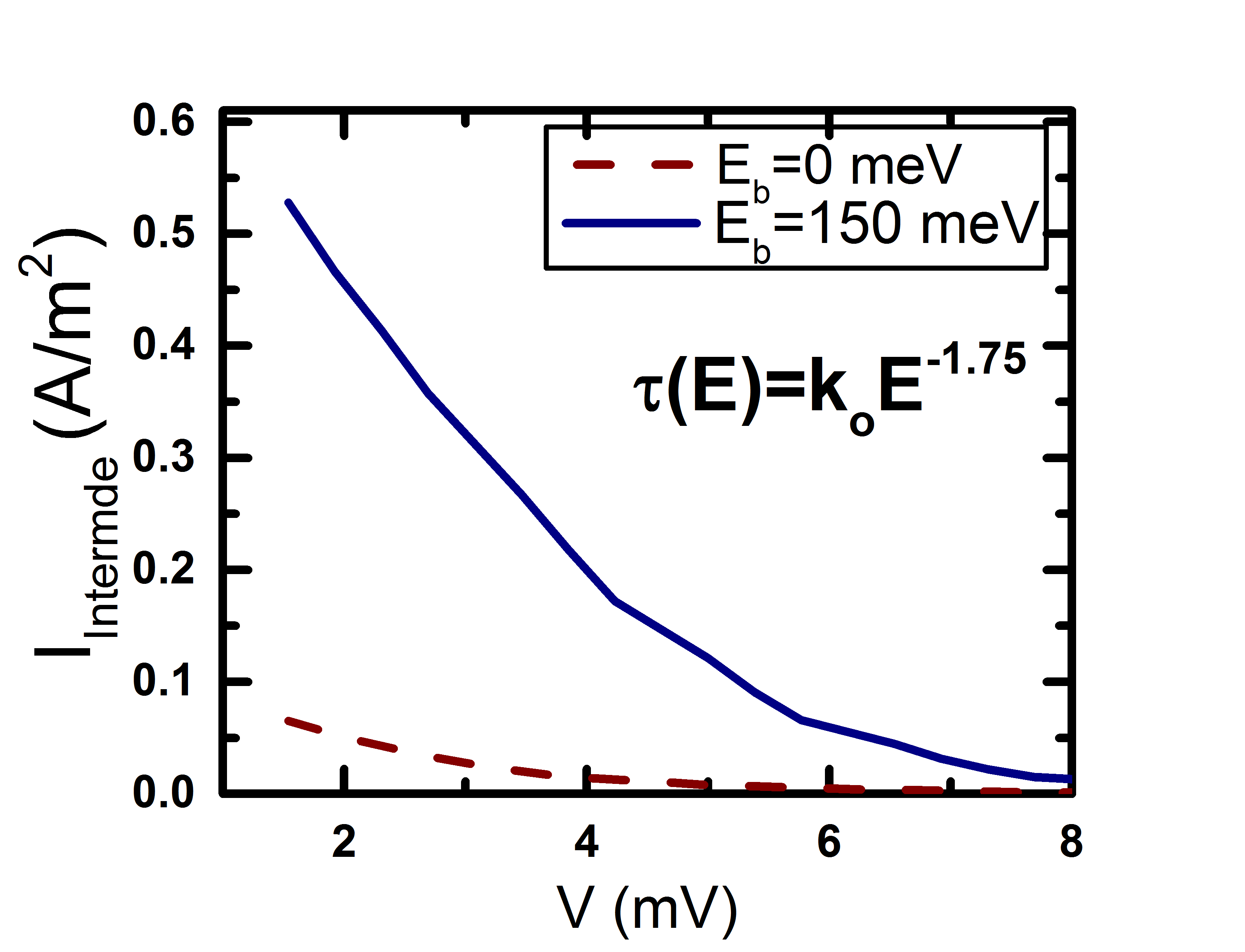}}

\caption{Plot of (a) $I_{Total}-I_{Intermode}$ at maximum power at a given voltage for  $r=-1.25$, (b) $I_{Intermode}$ at maximum power at a given voltage for $r=-1.25$, (c) $I_{Total}-I_{Intermode}$ at maximum power at a given voltage for   $r=-1.5$, (d) $I_{Intermode}$ at maximum power at a given voltage for $r=-1.5$, (e) $I_{Total}-I_{Intermode}$ at maximum power at a given voltage for   $r=-1.75$, and  (f) $I_{Intermode}$ at maximum power at a given voltage for $r=-1.75$. The value of $I_{Intermode}$ does not show a strong dependence on the value of $r$ because for moderate scattering rate, the flow of intermode coupling current depends on the number of empty states in the lower energy modes. Hence, a change in the rate of scattering does not drastically affect the intermode coupling current.  }
\label{fig:other_scat_intermode}
\end{figure}
We plot in Fig. \ref{fig:intermode1} the intermode current per unit energy per unit area profile of  bulk generators  at maximum power for various barrier heights    taking acoustic phonon scattering into account. It is shown that electronic current flows out (positive value) of the higher energy modes into (negative value) the lower energy modes. Such a flow of electronic current from the higher energy to the lower energy modes occurs in an attempt to restore quasi-equilibrium in the lower energy modes. The total current flow without the intermode current at the maximum power at a given voltage is shown in Fig. \ref{fig:intermode2}(a).  The overlapping curves in Fig. \ref{fig:intermode2}(a) demonstrate  that an  increase in the  flow of intermode current is the main reason for the  enhancement in generated power with an increase in  the height of the energy barrier  for acoustic phonon scattering. The plot of  intermode current at the maximum power at a given voltage is shown in  Fig \ref{fig:intermode2} (b) for three different barrier heights. The intermode current profile per unit area for other scattering mechanisms is shown in Fig. \ref{fig:other_scat_intermode}. The intermode current does not show a strong dependence on the value of $r$. This is because for moderate  scattering rate, the maximum intermode coupling current is  limited by the difference in electron population between the higher and the lower energy modes. The decrease in filtering coefficient with decrease in $r$ is mainly due to the  decrease in the current that directly flows from one contact to the other without any intermode transition. We hence conclude that electron-phonon scattering in bulk thermoelectric generators enhances the effect of energy filtering.
\subsection{Variation with device length}\label{length}
\begin{figure}
\includegraphics[scale=.3]{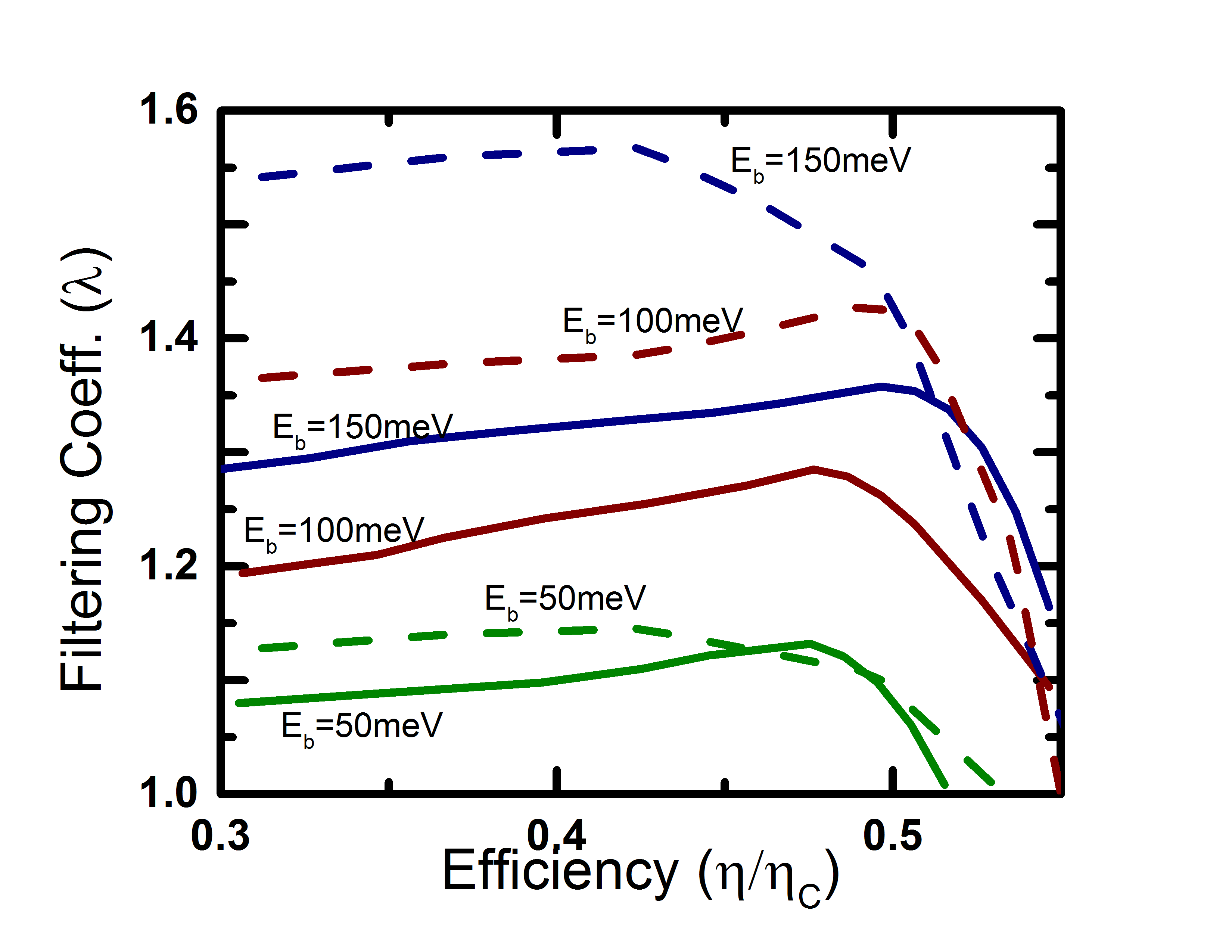}
\caption{Plot of the filtering coefficient ($\lambda$) versus efficiency ($\eta/\eta_C$)  for  various barrier heights for $2.85nm$ wide square nanowires of length $20nm$ (solid curves) and $40nm$ (dashed curves). The nanowires are  embedded with a Gaussian energy barrier ($\sigma_w=2.7nm$). }
\label{fig:variation_with_length}
\end{figure}
Manipulation of electron-phonon scattering is the key to an enhancement of generated power due to energy filtering. For nano-wires embedded with an energy barrier, scattering is most dominant near the top of the barrier where the density of states is maximum. For such cases, the true benefits of energy filtering can be harnessed when the device region  is much longer compared to the width of the energy barrier such that the electrons can undergo transport with very less scattering in the region where kinetic energy is very high. When the length of the energy barrier is made comparable to the length of the device, a huge amount of  scattering near the top of the barrier deteriorates the generated power. The rate of  electron-phonon scattering in bulk generators increases with energy. However, the increase in velocity of electrons along the transport direction with energy  mitigates the effect of this  increase in acoustic phonon scattering rate. In addition, it can be shown that the intermode current due to intermode coupling in bulk generators increase to its maximum when the  device region between the source contact and the barrier interface   is  longer that a few momentum relaxation lengths ($\lambda_p$). Such an increase in current also results in an  increase in the generated power. Plots of the filtering coefficient versus efficiency for nanowire thermoelectric generators  of length $20nm$ and $40nm$ are shown in Fig. \ref{fig:variation_with_length} for the same Gaussian energy barrier width ($\sigma_w=2.7nm$). 
\subsection{Perfect versus imperfect energy filtering for power generation}\label{imperfect}
Theoretical study of energy filtering in thermoelectric generators has demonstrated a maximum enhancement of power generation when energy filtering is perfect  \cite{theory1,theory3,whitney}. In ballistic conductors a sharp cutoff energy can easily be achieved with the help of a wide barrier \cite{theory3} or multiple resonant tunnel structures \cite{theory1}. However, for systems dominated by incoherent scattering, achieving a sharp cut-off energy for thermoelectric power generation may be quite  challenging. Incoherent scattering due phonon broadens the energy levels and creates local density of states such that perfect energy filtering is never achieved. %In addition to that a sharp cut-off energy requires wider barriers which might have finite size effects sharp decrease in the term $v_z^2(E)\tau (E)D(E)$ near the top of the barrier.
\begin{figure}[!htb]
\subfigure{\includegraphics[scale=.32]{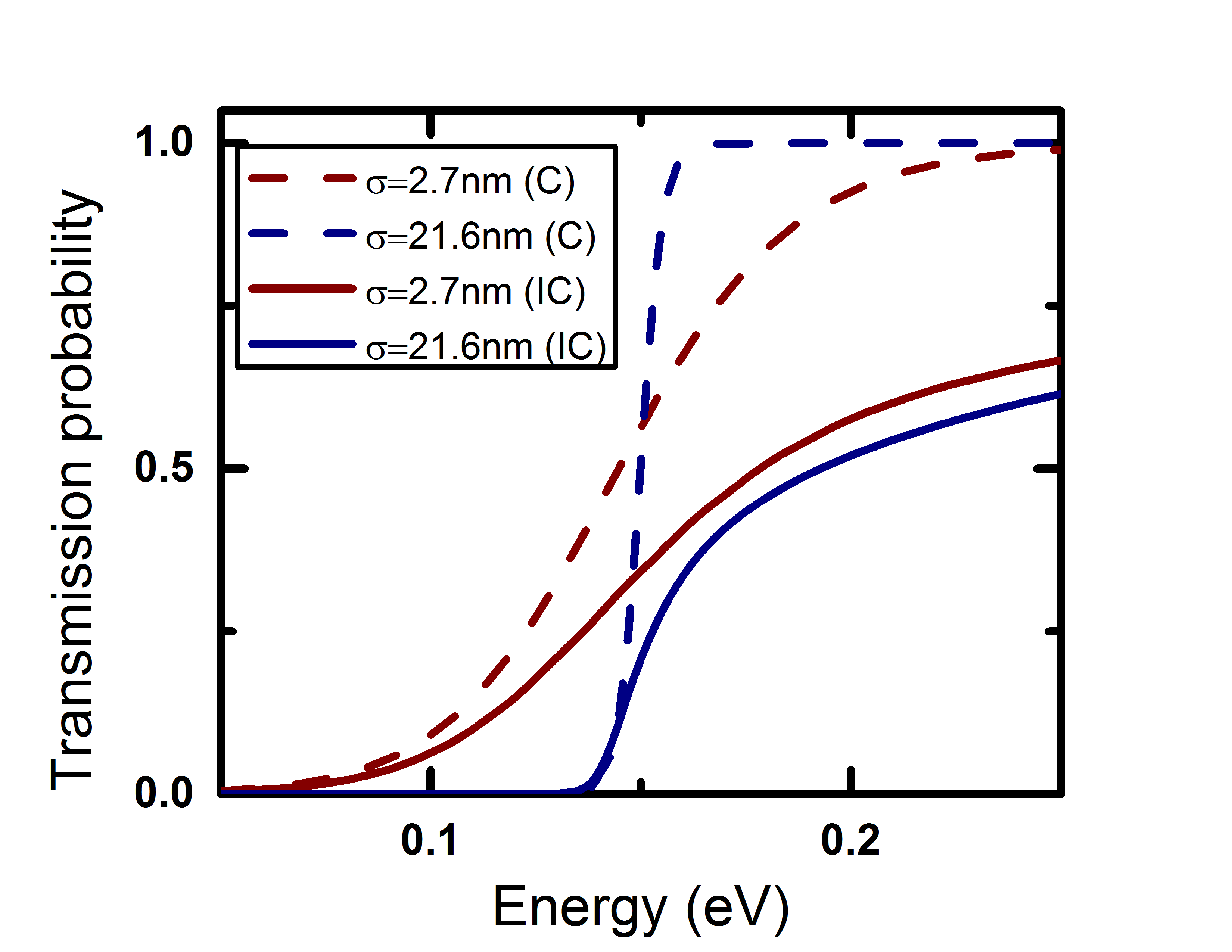}
}
\subfigure{\includegraphics[scale=.32]{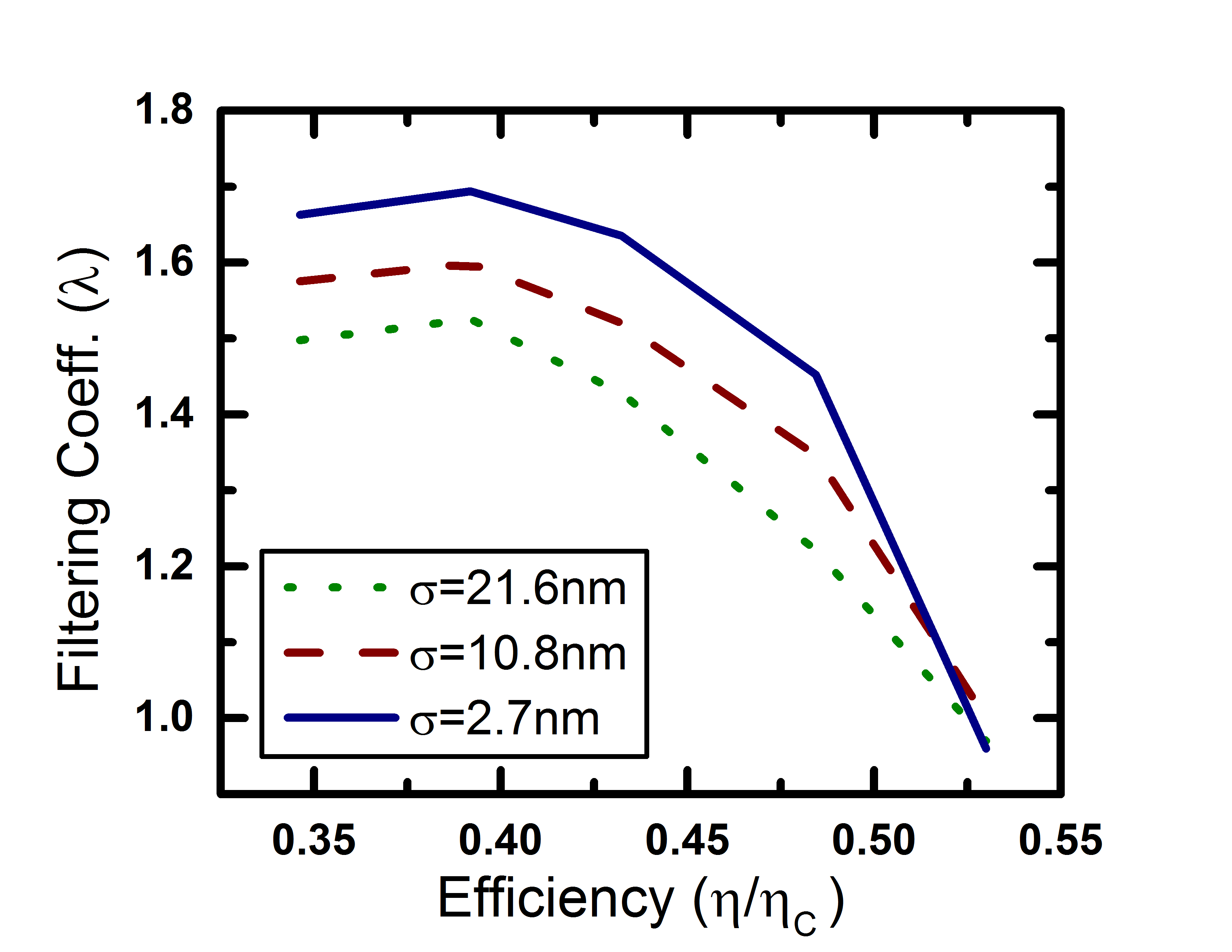}
}
\caption{Plot of (a) transmission probability of electrons versus energy for   Gaussian energy barriers of height $E_b=150meV$ and $\sigma_w=2.7nm$ and $~21.6nm$ for coherent (dashed lines) and incoherent  scattering (solid lines) and (b)  filtering coefficient versus efficiency ($\eta/\eta_C$) for  a $170nm$ long diffusive nanowire with different energy  barrier widths. Unlike the case of coherent transport (Fig. \ref{fig:ballistic_nanowire}), a wider barrier does not always lead to an increase in filtering coefficient($\lambda$) for devices dominated by incoherent scattering }
\label{fig:transmission}
\end{figure}
The transmission probability of electrons in nanowire generators dominated by coherent and incoherent  scattering is shown in Fig \ref{fig:transmission}(a). The thick barrier provides a sharper transmission  cut-off for ballistic devices and  hence is more suitable for power generation. However, for devices dominated by incoherent scattering, although a wider barrier provides a sharper  cut-off energy, the overall transmission probability through such  barriers might decrease due to drastic increase in scattering rates near the top of the barrier resulting in a decreasing current and generated power. Hence, depending on the length of the generator, there is an optimum barrier width at a given efficiency of operation for which power generation is maximum. %Complete optimization of barrier width at various efficiency and for varying device length is tedious and will not be shown here.   
The filtering coefficient ($\lambda$) over a range of efficiency is  shown in  Fig.  \ref{fig:transmission}(b) for various barrier widths in case of  a  $170nm$ long square nanowire of width $2.85nm$  taking  incoherent acoustic phonon scattering into account. The filtering coefficient ($\lambda$) for the thinner barrier is higher compared to  that for wider barriers due to overall decrease in $T=\int T(E) \Big \{ -\frac{\partial f}{\partial E} \Big \}dE$ for wider barriers.
\subsection{Combination of various scattering mechanisms} \label{combo}
So far, we have discussed the general conditions under which energy filtering enhances generated power in nanoscale and bulk thermoelectric generators. To demonstrate such conditions we have assumed smooth Gaussian barriers. However our discussion is valid for other types of barriers as well. In fabricated devices, a number of electron scattering mechanisms may be dominant such that the relaxation time of the electrons is a polynomial function of kinetic energy ($E$). 
\[
\tau(E)=\sum_i k_iE^{r_i}
\]
We split the contributions $r_i$ into two groups (a) $r_i\geq r_{min}$ (b) $r_i<r_{min}$. With energy filtering, thermoelectric power generation is enhanced in the presence of the scattering mechanisms satisfying $r_i\geq r_{min}$  while the same is degraded in the presence of the scattering mechanisms satisfying $r_i<r_{min}$. In our simulations, we have assumed a single scattering mechanism to be the dominant one such that $\tau(E)= k_iE^{r_i}$. If scattering mechanisms of both categories are present,  there is an optimum cut-off energy at which enhancement of generated power due to energy filtering is maximum \cite{theory1}. Also the change in effective mass due to non-parabolicity of the band-structure can have an effect on the scattering process. In addition to that, in degenerately doped semiconductors, scattering of electrons from a primary to a satellite valley may have drastic effect on enhancement of generated power due to decrease in kinetic energy (velocity) of the electrons in the satellite valley. All these considerations hamper the accurate theoretical prediction of the filtering coefficient in a specific  semiconductor. 
 \section{Conclusion} \label{conc}
In this paper, we have shown that the  enhancement in generated power due to energy filtering via planar energy barriers is a characteristic of devices dominated by incoherent scattering. In addition to discussing the role of the  parameter $\Upsilon=v^2_z(E)\tau(E)D(E)$ which determines the relative enhancement in generated power in single moded nanowires, we have demonstrated the effect of finite width of the barrier on the generated power as well as the role of intermode scattering on power generation in bulk thermoelectric generators. We have shown that electron-phonon scattering enhances the effect of energy filtering on generated power by driving electron population from the higher to the lower sub-bands.
In our paper we have assumed the length of the device to be less than the energy relaxation length ($L_E$). In case of longer devices, multiple energy barriers may be embedded within the same device keeping the distance between the barriers ($d_m$) maximum while maintaining $d_m<L_E$ \cite{kim2,kim1}.  In all our calculations we have neglected the decrease in efficiency due to phonon heat conductivity. Although the performance of both nanowire and bulk generators are effected due to phonon heat conductivity,  embedding energy barriers in nanowire/bulk provide  advantage in terms of decrease in lattice heat conductivity due to scattering of long wavelength phonons near the barrier  \cite{phonon1,phonon2,phonon3,phonon4,phonon5,phonon6,phonon7,superlattice1,superlattice2} or nanocomposite \cite{nanocomposite} interface.  Analysis of the results obtained in this work should provide general design guidelines for enhancement in power generation with energy filtering.

{\it{Acknowledgements:}} This work was partly supported by the IIT Bombay SEED grant and the Indian Space Research Organization RESPOND grant. The authors acknowledge Gang Chen for valuable suggestions and S. D. Mahanti for insightful discussions.
\appendix
\section{NEGF equations for intermode coupling}\label{appendix1}
In case of non-dissipative transport in nano devices, the generalized equations  for Green's function  and scattering matrix are  given by:
\begin{gather} 
G(\overrightarrow{k_{m}},E_z)=[E_zI-H-U-\Sigma(\overrightarrow{k_{m}},E_z)]^{-1} \nonumber \\
\Sigma(\overrightarrow{k_{m}},E_z)=\Sigma_L(\overrightarrow{k_{m}},E_z)+\Sigma_R(\overrightarrow{k_{m}},E_z)+\Sigma_s(\overrightarrow{k_{m}},E_z) \nonumber \\
 A(\overrightarrow{k_{m}},E_z)=i[G(\overrightarrow{k_{m}},E_z)-G^{\dagger}(\overrightarrow{k_{m}},E_z)] \nonumber \\
 \Gamma(\overrightarrow{k_{m}},E_z)=[\Sigma(\overrightarrow{k_{m}},E_z)-\Sigma^{\dagger}(\overrightarrow{k_{m}},E_z)], \nonumber \\
 \label{eq:negf}  
\end{gather}

where $H$ is the discretized Hamiltonian matrix (constructed using  the effective mass approach \cite{dattabook}), $U=-qV$ is the electronic potential energy in the band and $\Sigma_L(\overrightarrow{k_{m}},E_z)+\Sigma_R(\overrightarrow{k_{m}},E_z)$ and $\Sigma_s(\overrightarrow{k_{m}},E_z)$ describe the effect of  coupling and scattering of electronic wavefunctions  due to contacts and electron-phonon interaction respectively. In the above sets of equations, $\overrightarrow{k_{m}}$ denote the transverse wave-vector and $E_z$ is the free variable denoting the energy of the electrons  along the transport direction. $A(\overrightarrow{k_{m}},E_z)$ is the $1-D$ spectral function  for the $m^{th}$ sub-band and $\Gamma(\overrightarrow{k_{m}},E_z)$ is the broadening matrix for the $m^{th}$ sub-band at longitudinal energy $E_z$. 
For moderate electron-phonon interaction, it is generally assumed that the real part of $\Sigma_s=0$. Hence, 
\begin{equation}
\Sigma_s(\overrightarrow{k_{m}},E_z)=i\frac{ \Gamma_s(\overrightarrow{k_{m}},E_z)}{2}=\Sigma^{in}_s(\overrightarrow{k_{m}},E_z)+\Sigma^{out}_s(\overrightarrow{k_{m}},E_z)
\label{eq:sigma_phonon}
\end{equation}
$ \Sigma^{in}(\overrightarrow{k_{m}},E_z)$ and $ \Sigma^{out}(\overrightarrow{k_{m}},E_z)$ are the inscattering and the  outscattering functions which models the rate of scattering of the  electrons from the contact and inside the device.

\begin{multline}
  \Sigma^{in}(\overrightarrow{k_{m}},E_z)=\Sigma^{in}_L(\overrightarrow{k_{m}},E_z)+\Sigma^{in}_R(\overrightarrow{k_{m}},E_z) \\
  +\Sigma^{in}_s(\overrightarrow{k_{m}},E_z) \nonumber 
  \end{multline}
  \begin{multline}
   \Sigma^{out}(\overrightarrow{k_{m}},E_z)=\Sigma^{out}_L(\overrightarrow{k_{m}},E_z)+\Sigma^{out}_R(\overrightarrow{k_{m}},E_z) \\
   +\Sigma^{out}_s(\overrightarrow{k_{m}},E_z) \nonumber 
   \label{eq:sig}
  \end{multline}
   where the subscript $'L'$, $'R'$ and $'s'$ denote the influence of left contact, right contact and electron-phonon scattering respectively. The in-scattering and out-scattering functions are dependent on the contact quasi-Fermi distribution functions as:
   
\begin{gather}
 \Sigma^{in}(\overrightarrow{k_{m}},E_z)=\underbrace{\Gamma_L(E_z)f_L(E_z+\frac{\hslash^2 \overrightarrow{k_m}^2}{2m_t})}_{inflow~from~left~contact} \nonumber \\+\underbrace{\Gamma_R(E_z)f_R(E_z+\frac{\hslash^2 \overrightarrow{k_m}^2}{2m_t})}_{inflow~from~right~contact}+\underbrace{\Sigma^{in}_s(\overrightarrow{k_{m}},E_z)}_{inflow~due~to~phonons} \nonumber  \\
 \Sigma^{out}(\overrightarrow{k_{m}},E_z)=\underbrace{\Gamma_L(E_z)\Big\{ 1-f_L(E_z+\frac{\hslash^2 \overrightarrow{k_m}^2}{2m_t})\Big \} }_{outflow~to~left~contact} \nonumber \\+\underbrace{\Gamma_R(E_z)\Big \{ 1-f_R(E_z+\frac{\hslash^2 \overrightarrow{k_m}^2}{2m_t})\Big \}}_{outflow~to~right~contact}+\underbrace{\Sigma^{out}_s(\overrightarrow{k_{m}},E_z)}_{outflow~due~to~phonons} \nonumber  \\
 \label{eq:sig1}
 \end{gather}
where $f_{L(R)}$ denote the quasi-Fermi distribution of left(right) contact.
The rate of scattering of electrons due to phonons is dependent on the electron and the hole correlation functions $(G^n$ and $G^p)$ and is given by:

\begin{gather}
\Sigma^{in}_s(\overrightarrow{k_{m}},E_z)=D_O
\underset{\overrightarrow{q_{t}}}{\sum}G^n(\overrightarrow{k_{m}}+\overrightarrow{q_{t}},E_z-\Delta E_{\overrightarrow{k_{m}}+\overrightarrow{q_{t}},\overrightarrow{k_{m}}}) \nonumber \\
\Sigma^{out}_s(\overrightarrow{k_{m}},E_z)=D_O
\underset{\overrightarrow{q_{t}}}{\sum}G^p(\overrightarrow{k_{m}}+\overrightarrow{q_{t}},E_z-\Delta E_{\overrightarrow{k_{m}}+\overrightarrow{q_{t}},\overrightarrow{k_{m}}})
\label{eq:sigma}
\end{gather}

where $\Delta E_{\overrightarrow{k_{m}}+\overrightarrow{q_{t}},\overrightarrow{k_{m}}}=\frac{\hslash^2 (\overrightarrow{k_{m}}+\overrightarrow{q_{t}})^{2}}{2m_t}-\frac{\hslash^2 \overrightarrow{k_{m}}^{2}}{2m_t}$.

 $G^n(\overrightarrow{k_{m}},E_z)$ and $G^p(\overrightarrow{k_{m}},E_z)$ are the electron and the hole correlation functions for the $m^{th}$ sub-band and $\{\overrightarrow{q_t}\}$ denotes the set of transverse phonon wave vectors. The electron and the hole correlation functions are again related to the  electron in-scattering and the electron out-scattering functions via the equations:
\begin{gather}
G^n(\overrightarrow{k_{m}},E_z)=G(\overrightarrow{k_{m}},E_z)\Sigma^{in}(\overrightarrow{k_{m}},E_z)G^{\dagger}(\overrightarrow{k_{m}},E_z) \nonumber \\
G^p(\overrightarrow{k_{m}},E_z)=G(\overrightarrow{k_{m}},E_z)\Sigma^{out}(\overrightarrow{k_{m}},E_z)G^{\dagger}(\overrightarrow{k_{m}},E_z) \nonumber \\
 \label{eq:correlation} 
\end{gather}
 Solving the  dynamics of the entire system involves a self consistent solution of \eqref{eq:negf},  \eqref{eq:sigma_phonon}, \eqref{eq:sigma} and  \eqref{eq:correlation}.
 For momentum scattering due to acoustic phonons, $D_O$ in the above equations can be related to the acoustic phonon deformation potential ($D_{ac}$) by:

\begin{equation}
D_O=\frac{ D_{ac}^2k_BTF}{\rho v_s^2 a^3}
\end{equation}

where $F$ is known as the form factor and denotes the spacial spread of the phonon wave-vectors. $\rho$ and $v_s$ denote the mass density and the velocity of sound in the material respectively. For the purpose of our simulation, we have used the parameters of bulk silicon.

  The spectral function for the $m^{th}$ sub-band is given by 
\[
A(\overrightarrow{k_{m}},E_z)=G^n(\overrightarrow{k_{m}},E_z)+G^p(\overrightarrow{k_{m}},E_z)
\]

The electron density and current at the grid point $j$ can be calculated from the above equations as:

\[
n_j=\underset{m}{\sum}\int\frac{[G^n(\overrightarrow{k_{m}},E_z)dE_z]}{\pi aA}
\]

\begin{eqnarray}
I^{j\rightarrow j+1}=\underset{k_m}{\sum}\frac{q}{\pi \hslash} Im\int[H_{j+1,j}(E_z)G^n_{j,j+1}(\overrightarrow{k_{m}},E_z) \nonumber \\
-G^n_{j+1,j}(\overrightarrow{k_{m}},E_z)H_{j,j+1}(E_z)]dE_z \nonumber \\
\label{eq:currentnegf}
\end{eqnarray}

where $a$ is the distance between two adjacent grid points and $A$ is the cross sectional area of the device. $\hslash k_m$ denotes the transverse momentum of the electrons in the $m^{th}$ sub-band. The summations in \eqref{eq:currentnegf} run over all the sub-bands available for conduction.

The heat current flowing through the device is given by:

\begin{eqnarray}
I_Q^{j\rightarrow j+1}=\underset{k_m}{\sum}\frac{1}{\pi \hslash} \times (E_z+E_m-\mu_H) Im\int[H_{j+1,j}(E_z)\nonumber \\
G^n_{j,j+1}(\overrightarrow{k_{m}},E_z) 
-G^n_{j+1,j}(\overrightarrow{k_{m}},E_z)H_{j,j+1}(E_z)]dE_z \nonumber \\
\label{eq:heatcurrentnegf}
\end{eqnarray}

where $E_z$ is the kinetic energy of the electrons due to momentum along the transport direction and  $E_m$ is the kinetic energy of the electron due to momentum in the transverse direction. 

\section{Approximate derivation of scattering self energies for higher order scattering mechanisms} \label{appendix2}

For elastic scattering, the Boltzmann transport equation is given by: \cite{lundstrombook,book1}: 
\begin{eqnarray}
\frac{\partial f(r,\overrightarrow{k},t)}{\partial t}=\sum_{\overrightarrow{k'}}\Big\{\underbrace{S(\overrightarrow{k'},\overrightarrow{k})\{1-f(r,\overrightarrow{k},t)\}  f(r,\overrightarrow{k'},t)}_{in-scattering} \nonumber\\
 -\underbrace{S(\overrightarrow{k},\overrightarrow{k'})\{1-f(r,\overrightarrow{k'},t)\}f(r,\overrightarrow{k},t)}_{out-scattering}\Big\} \delta(E_k-E_{k'}) \nonumber \\ 
 \label{eq:boltzmann}
\end{eqnarray}
$S(\overrightarrow{k},\overrightarrow{k'})/S(\overrightarrow{k'},\overrightarrow{k})$ incorporate the dependence of the rate of electron scattering on energy/momentum. For isotropic scattering with acoustic phonons, $S(\overrightarrow{k'},\overrightarrow{k})$ is independent of $\overrightarrow{k'}$ or $\overrightarrow{k}$
\begin{eqnarray}
S(\overrightarrow{k'},\overrightarrow{k})=S(\overrightarrow{k},\overrightarrow{k'})=S(E_{\overrightarrow{k}})=S(E_{\overrightarrow{k'}})\nonumber \\
=\frac{2\pi k_BTD_{ac}^2}{\rho \hslash v_s^2A}
\label{eq:s}
\end{eqnarray}
where   $D_{ac}$, $\rho$ and $v_s$ are the acoustic deformation potential, the mass density and the velocity of sound in the medium respectively  \cite{lundstrombook,book1}. The right side of Eq. \ref{eq:boltzmann} can be simplified by summing over the states  (assuming steady state) \cite{lundstrombook,book1}:

\begin{eqnarray}
\frac{\partial f(r,\overrightarrow{k},t)}{\partial t}=\{1-f(r,\overrightarrow{k})\}\sum_{\overrightarrow{k'}}S(\overrightarrow{k'},\overrightarrow{k})  f(r,\overrightarrow{k'}) \delta(E_k-E_{k'}) \nonumber\\
 -f(r,\overrightarrow{k})\sum_{\overrightarrow{k'}}S(\overrightarrow{k},\overrightarrow{k'})\{1-f(r,\overrightarrow{k'})\} \delta(E_k-E_{k'}) \nonumber \\ 
 =\{1-f(r,\overrightarrow{k})\}S(E_{\overrightarrow{k}})\sum_{\overrightarrow{k'}}  f(r,\overrightarrow{k'}) \delta(E_k-E_{k'}) \nonumber\\
 -f(r,\overrightarrow{k})S(E_{\overrightarrow{k}})\sum_{\overrightarrow{k'}}\{1-f(r,\overrightarrow{k'})\} \delta(E_k-E_{k'}) \nonumber \\ 
\end{eqnarray}

%Putting $E_{\overrightarrow{k}}=E_{\overrightarrow{k'}}=E$, we get
\begin{eqnarray}
\Rightarrow \frac{\partial f(r,\overrightarrow{k})}{\partial t}
 =\{1-f(r,\overrightarrow{k})\}\underbrace{S(E_{\overrightarrow{k}})n_{tot}(r,E_{\overrightarrow{k}})}_{\frac{2\pi}{\hslash}\Sigma^{in}(E_{\overrightarrow{k}})} \nonumber \\
 -f(r,\overrightarrow{k})\underbrace{S(E_{\overrightarrow{k}})p_{tot}(r,E_{\overrightarrow{k}})}_{\frac{2 \pi}{\hslash}\Sigma^{out}(E_{\overrightarrow{k}})} \nonumber \\ 
\end{eqnarray}
where 
\begin{gather}
n_{tot}(r,E_{\overrightarrow{k}})=\sum_{\overrightarrow{k'}}n(r,E_{\overrightarrow{k'}})\delta(E_{\overrightarrow{k}}-E_{\overrightarrow{k'}}) \nonumber \\
 p_{tot}(r,E_{\overrightarrow{k}})=\sum_{\overrightarrow{k'}}p(r,E_{\overrightarrow{k'}})\delta(E_{\overrightarrow{k}}-E_{\overrightarrow{k'}}) \nonumber 
\end{gather}
For acoustic phonon, $S(E_{\overrightarrow{k}})$ is independent of $E_{\overrightarrow{k}}$, $n_{tot}(E_{\overrightarrow{k}}) \approx D(E)f(r,E) $ and  $p_{tot}(E_{\overrightarrow{k}}) \approx D(E)\{1-f(r,E)\} $. Therefore, 
\begin{gather}
\tau(E_{\overrightarrow{k}}) \propto \frac{1}{\frac{\partial f(r,\overrightarrow{k})}{\partial t}} \propto \frac{1}{D(E)} \nonumber \\
\Rightarrow \tau(E_{\overrightarrow{k}}) \propto E_{\overrightarrow{k}}^n
\end{gather}

where $n=0.5,~0,~-0.5$ for $1-D,~2-D$ and $3-D$ devices respectively. To demonstrate the effect of the   scattering which are of order higher than  phonon scattering, we choose 
\[
S(E_{\overrightarrow{k}})=kE_{\overrightarrow{k}}^r
\]
where $k$ is a constant of proportionality, such that 
\[
\tau(E_{\overrightarrow{k}}) \propto E_{\overrightarrow{k}}^{n+r}
\]
where $n$ is same as defined above. In NEGF, we then use
\begin{gather}
\Sigma^{in}(r,E_{\overrightarrow{k}})=\frac{\hslash}{2\pi} S(E_{\overrightarrow{k}})\sum_{\overrightarrow{k'}}n(r,E_{\overrightarrow{k'}})\delta(E_{\overrightarrow{k}}-E_{\overrightarrow{k'}}) \nonumber \\
\Sigma^{out}(r,E_{\overrightarrow{k}})=\frac{\hslash}{2\pi} S(E_{\overrightarrow{k}})\sum_{\overrightarrow{k'}}p(r,E_{\overrightarrow{k'}})\delta(E_{\overrightarrow{k}}-E_{\overrightarrow{k'}}) \nonumber \\
\end{gather}

\section{Derivation of the factor $\Upsilon$}\label{appendix3}
In case of diffusive or incoherent transport without externally applied magnetic field, the dynamics of the electron system follows the quasi-distribution function given by \cite{ashcroft}

\begin{eqnarray}
f(\overrightarrow{k})=f_0(E_{\overrightarrow{k}})+\int^{\infty}_{0}P(\overrightarrow{k},\tau')\Big\{(-\frac{\partial f_0}{\partial E}) \overrightarrow{v}(\overrightarrow{k}).\Big(-e\overrightarrow{\mathlarger{\mathlarger{\mathlarger{\varepsilon}}}} \nonumber \\
- \nabla \mu -\frac{E-\mu}{T} \nabla T\Big)\Big\}d\tau' \nonumber \\
\label{eq:diffusive_f}
\end{eqnarray}

where $P(\overrightarrow{k},\tau')$ is the fraction of the electrons with wavevector $\overrightarrow{k}$ that donot suffer a scattering within the time period $\tau'$. For isotropic scattering, generally $P(\overrightarrow{k},\tau')$ takes the form \cite{ashcroft}:

\begin{equation}
P(\overrightarrow{k},\tau')=e^{\frac{-\tau'}{\tau(\overrightarrow{k})}}
\end{equation}
Generally for isotropic and local scattering processes, $\tau(\overrightarrow{k})$ depends on $\overrightarrow{k}$ through the energy $E_{\overrightarrow{k}}$. Therefore, 

\begin{equation}
P(\overrightarrow{k},\tau')=e^{{-\tau'}/{\tau(E_{\overrightarrow{k}})}}
\end{equation}

Equation \eqref{eq:diffusive_f} then becomes	
\begin{eqnarray}
f(\overrightarrow{k})=f_0(E_{\overrightarrow{k}})+\int^{\infty}_{0}e^{{-\tau'}/{\tau(E_{\overrightarrow{k}})}}\Big\{(-\frac{\partial f_0}{\partial E}) \overrightarrow{v}(\overrightarrow{k}).\Big(-e\overrightarrow{\mathlarger{\mathlarger{\mathlarger{\varepsilon}}}} \nonumber \\
- \nabla \mu -\frac{E-\mu}{T} \nabla T\Big)\Big\}d\tau' \nonumber \\
\Rightarrow f(\overrightarrow{k})=f_0(E_{\overrightarrow{k}})+\tau(E_{\overrightarrow{k}}) \overrightarrow{v}(\overrightarrow{k}).\Big\{(-\frac{\partial f_0}{\partial E})\Big(-e\overrightarrow{\mathlarger{\mathlarger{\mathlarger{\varepsilon}}}} 
- \nabla \mu \nonumber \\-\frac{E-\mu}{T} \nabla T\Big)\Big\} \nonumber \\
\end{eqnarray}

Assuming that the potential and the temperature gradient are applied in the $z$ direction only,
\begin{eqnarray}
f(\overrightarrow{k})=f_0(E_{\overrightarrow{k}})+\tau(E_{\overrightarrow{k}}) {v_z}(\overrightarrow{k})\Big\{(-\frac{\partial f_0}{\partial E})\Big(-e\overrightarrow{\mathlarger{\mathlarger{\mathlarger{\varepsilon_z}}}} 
-\frac{\partial \mu(z)}{\partial z} \nonumber \\-\frac{E-\mu(z)}{T(z)} \frac{\partial T(z)}{\partial z} \Big)\Big\} \nonumber \\
\end{eqnarray}

The current density in the z-direction, is therefore, given by:
\[
{j_z}=-e\int \frac{d\overrightarrow{k}}{4 \pi^3}v_z(\overrightarrow{k})f(\overrightarrow{k}) 
\]
\begin{eqnarray}
=-e\int \frac{d\overrightarrow{k}}{4 \pi^3}v_z(\overrightarrow{k}) \Big[ f_0(E_{\overrightarrow{k}})+\tau(E_{\overrightarrow{k}}) {v_z}(\overrightarrow{k})\Big\{(-\frac{\partial f_0}{\partial E}) \nonumber \\ \Big(-e\overrightarrow{\mathlarger{\mathlarger{\mathlarger{\varepsilon_z}}}} 
-\frac{\partial \mu(z)}{\partial z}-\frac{E-\mu(z)}{T(z)} \frac{\partial T(z)}{\partial z} \Big)\Big\}\Big] \nonumber \\
\end{eqnarray}

The integral of the term $v_z(\overrightarrow{k}) f_0(E_{\overrightarrow{k}})$ vanishes since $f_0$ depends only on energy and is symmetrical in $\overrightarrow{k}$ space.

\begin{eqnarray}
j_z=-e\int \frac{d\overrightarrow{k}}{4 \pi^3} \tau(E_{\overrightarrow{k}}) \|\overrightarrow{v_z}(\overrightarrow{k})\|^2\Big\{(-\frac{\partial f_0}{\partial E})\Big(-e\overrightarrow{\mathlarger{\mathlarger{\mathlarger{\varepsilon_z}}}}  \nonumber \\
-\frac{\partial \mu(z)}{\partial z}-\frac{E-\mu(z)}{T(z)} \frac{\partial T(z)}{\partial z} \Big)\Big\} \nonumber \\
\label{eq:final_curr}
\end{eqnarray}

$\tau(E_{\overrightarrow{k}})$ and $\|\overrightarrow{v_z}(\overrightarrow{k})\|^2$ depend on $\overrightarrow{k}$ only through the energy $E_{\overrightarrow{k}}$. We can simplify \eqref{eq:final_curr} to transform $\overrightarrow{k}$ dependence to energy $(E)$ dependence:

\begin{eqnarray}
j_z=-e\int \tau(E) \|\overrightarrow{v_z}(E)\|^2 D(E)\Big\{(-\frac{\partial f_0}{\partial E})\Big(-e\overrightarrow{\mathlarger{\mathlarger{\mathlarger{\varepsilon_z}}}}  \nonumber \\
-\frac{\partial \mu(z)}{\partial z}-\frac{E-\mu(z)}{T(z)} \frac{\partial T(z)}{\partial z} \Big)\Big\}dE \nonumber \\
\end{eqnarray}

The term within the second bracket is the driving force for  the current and the term $\tau(E) \|\overrightarrow{v_z}(E)\|^2 D(E)$ defines the ease with which the driving force can cause a flow of the current. For the same applied voltage and temperature gradient (assuming that the driving force is same for devices with and without energy barriers), the generated power would increase with increase in the current. In case of perfect filtering, assuming that the length of the device is much greater than the length of the barrier (such that scattering near the top of the barrier does not affect the generated power appreciably),  a sufficient but not necessary condition for improvement of generated power with filtering is that $\tau(E) \|\overrightarrow{v_z}(E)\|^2D(E)$ is an increasing function of $E$. In other words,
\begin{equation}
\frac{\tau(E+E_b) \|\overrightarrow{v_z}(E+E_b)\|^2D(E+E_b)}{\tau(E) \|\overrightarrow{v_z}(E)\|^2D(E)}>1
\label{eq:condition}
\end{equation}

for $E_b>0$. Here $E_b$ is the cut-off energy for filtering. For isotropic and local scattering processes, $\tau(E)$ can generally be approximated as $\tau(E)=\sum_i k_iE^{r_i}$. In case of single moded nanowires, $\|\overrightarrow{v_z}(E)\|^2D(E) =2\sqrt{\frac{2\pi E}{m_lh^2}} $. The minimum value of $r$ for which energy filtering can enhance the generated power  in case of perfect filtering is therefore $r>r_{min}=-\frac{1}{2}$. For imperfect filtering, the value of $r_{min}$ may further increase.

For bulk generators, the value of $D(E)$ contributing to conduction cannot be defined properly due to partial momentum conservation. However assuming uncoupled mode transport, we can draw an upper limit on the value of  $r_{min}$. It can be shown that for perfect filtering and no effect of scattering near the barrier on the performance of the device, an assumption of uncoupled modes in electron transport  gives:

\begin{gather}
<v_z^2(E)D(E)>=\sum_m v_z^2(E-E_m)D_{1D}(E-E_m)  \nonumber \\
 = \int_0^{(E-E_b)} 2\frac{(E-E_m)}{m_l} \sqrt{\frac{2\pi m_l}{h^2}}\frac{1}{\sqrt{E-E_m}} \left(\frac{4\pi m_t}{h^2}dE_m\right) \nonumber \\
 =\frac{16\pi}{3}\frac{1}{h^3}\sqrt{\frac{2\pi m_t^2}{m_l}} (E^{\frac{3}{2}}-E_b^{\frac{3}{2}} ) \nonumber \\  
\end{gather}

for $E>E_b$. Here $<>$ denotes the average value of the argument and $m$ denotes all possible modes that are available for conduction.

\begin{figure}[!htb]
\includegraphics[scale=.26]{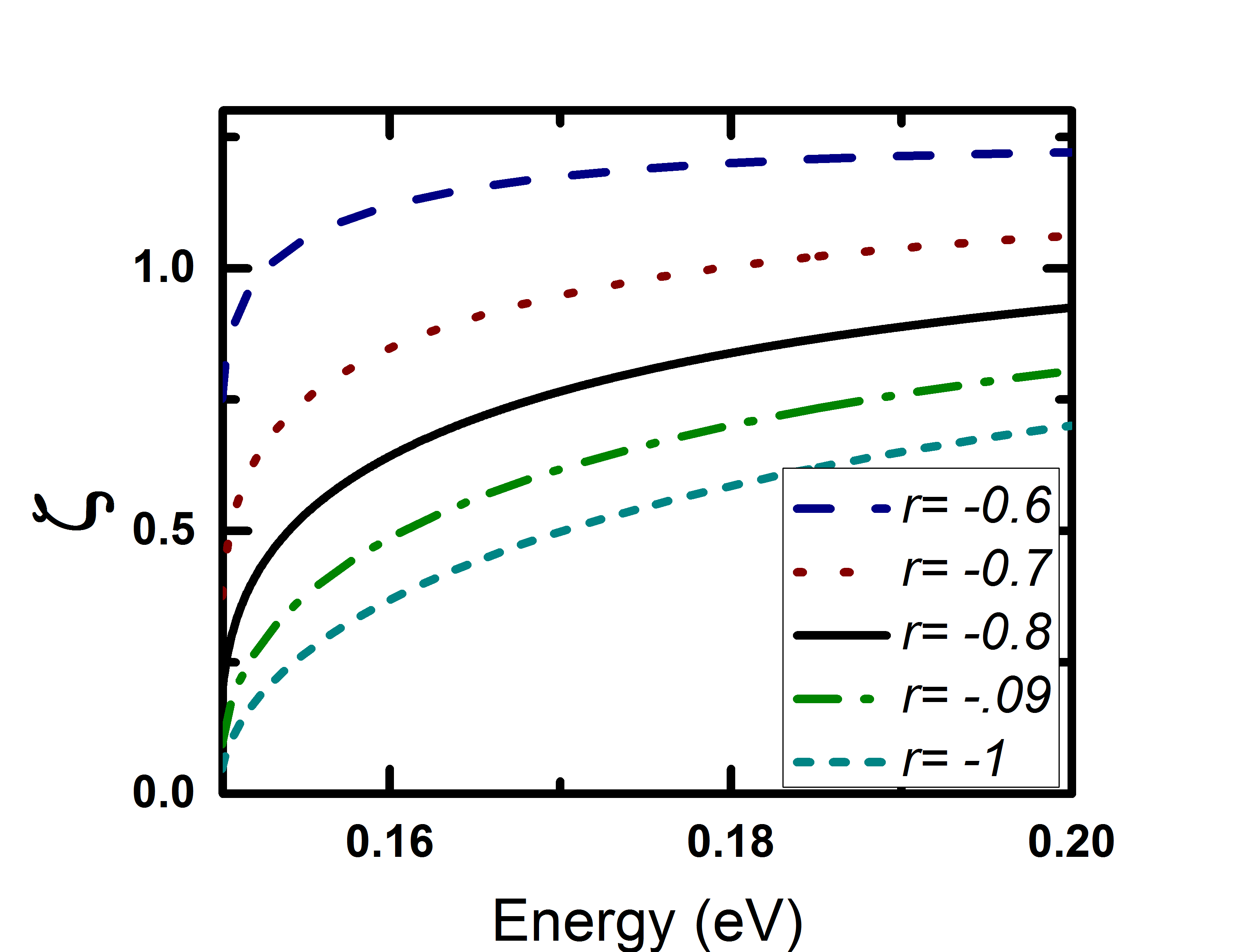}
\caption{Plot of the factor $\zeta=\frac{ (E^{\frac{3}{2}}-E_b^{\frac{3}{2}} )E^r}{ \{(E-E_b)^{r+\frac{3}{2}} \}}$ for various values of $r$ at $E_b=0.15eV$.}
\label{fig:zeta}
\end{figure}

 Assuming $\tau(E)=k_oE^r$, \eqref{eq:condition} translates to:
\[
\zeta=\frac{ (E^{\frac{3}{2}}-E_b^{\frac{3}{2}} )E^r}{ \{(E-E_b)^{r+\frac{3}{2}} \}}>1
\].
It can be shown that for $E_b=0.15eV$, the above condition is valid for $r>-0.6$ (See Fig. \ref{fig:zeta}).

\vfill

\bibliography{References}% Produces the bibliography via BibTeX.

\end{document}